\documentclass[useAMS,usenatbib]{mn2e}
\usepackage{graphicx}
\usepackage{textcomp}
\usepackage{threeparttable}
\usepackage{longtable}
\usepackage[fleqn]{amsmath}
\usepackage{lscape}
\usepackage{pdflscape}
\usepackage{dcolumn}
\usepackage{rccol}
\usepackage{environ}
\usepackage{threeparttablex}
\usepackage{amssymb}
\usepackage{comment} 

\newcommand{\mstellar}{\ensuremath{M_{\mathrm{stellar}}}}
\newcommand{\omatter}{\ensuremath{\Omega_{\mathrm{M}}}}

\newcommand{\col}{\ensuremath{\mathcal{C}}}

\newcommand{\msun}{\ensuremath{M_{\sun}}}

\newcommand{\nii}{\ensuremath{\mathrm{N}\,\textsc{ii}}}

\newcommand{\Caii}{\ensuremath{\mathrm{Ca}\,\textsc{ii}}}
\newcommand{\Caiii}{\ensuremath{\mathrm{Ca}\,\textsc{iii}}}
\newcommand{\Siii}{\ensuremath{\mathrm{Si}\,\textsc{ii}\,\lambda6355}}
\newcommand{\Siiihk}{\ensuremath{\mathrm{Si}\,\textsc{ii}\,\lambda3858}}
\newcommand{\Siiitmp}{\ensuremath{\mathrm{Si}\,\textsc{ii}\,\lambda5972}}
\newcommand{\vsiii}{\ensuremath{v_{\mathrm{Si}\,\textsc{ii}}}}
\newcommand{\vcaii}{\ensuremath{v_{\mathrm{Ca}\,\textsc{ii\,nir}}}}
\newcommand{\vcaiihk}{\ensuremath{v_{\mathrm{Ca}\,\textsc{ii\,H\&K}}}}

\newcommand{\rhvf}{\ensuremath{\mathrm{R_{HVF}}}}

\newcommand{\rgal}{\ensuremath{\mathrm{R}_{\mathrm{gal}}}}
\newcommand{\rsn}{\ensuremath{\mathrm{R}_{\mathrm{SN}}}}
\newcommand{\offset}{\ensuremath{\rsn/\rgal}}
\newcommand{\rptf}{\ensuremath{R_{\mathrm{P48}}}}
\newcommand{\gptf}{\ensuremath{g_{\mathrm{P48}}}}

\defcitealias{2004MNRAS.348L..59P}{PP04}
\defcitealias{2008ApJ...681.1183K}{KE08}
\defcitealias{2002ApJS..142...35K}{KD02}

\title[The host galaxies of SNe Ia]{Type Ia Supernova Spectral Features
in the Context of Their Host Galaxy Properties}
\author[Pan et al.]{Y.-C. Pan$^{1,2}$\thanks{E-mail:Yen-Chen.Pan@astro.ox.ac.uk}, M. Sullivan$^{2}$, K. Maguire${^3}$,
A. Gal-Yam$^{4}$, I. M. Hook$^{1,5}$,\newauthor D. A. Howell$^{6,7}$, P. E. Nugent$^{8,9}$, P. A. Mazzali$^{10,11,12}$\\
  $^{1}$Department of Physics (Astrophysics), University of Oxford, DWB, Keble Road, Oxford OX1 3RH, UK\\
  $^{2}$School of Physics and Astronomy, University of Southampton, Southampton, SO17 1BJ, UK\\
  $^{3}$European Southern Observatory (ESO), Karl-Schwarzchild-Str.2, 85748 Garching b. M\"unchen, Germany\\
  $^{4}$Department of Particle Physics and Astrophysics, Weizmann Institute of Science, Rehovot 76100, Israel\\
  $^{5}$INAF Osservatorio Astronomico di Roma, via Frascati, 33, 00040 Monte Porzio Catone, Roma, Italy\\
  $^{6}$Las Cumbres Observatory Global Telescope Network, Goleta, CA 93117, USA\\
  $^{7}$Department of Physics, University of California, Santa Barbara, CA 93106-9530, USA\\
  $^{8}$Department of Astronomy, University of California, Berkeley, CA 94720-3411, USA\\
  $^{9}$Computational Cosmology Center, Lawrence Berkeley National Laboratory, 1 Cyclotron Road, Berkeley, CA 94720, USA\\
  $^{10}$Astrophysics Research Institute, Liverpool John Moores University, 146 Brownlow Hill, Liverpool L3 5RF, UK\\
  $^{11}$INAF -- Osservatorio Astronomico, vicolo dellÕOsservatorio, 5, I-35122 Padova, Italy\\
  $^{12}$Max-Planck Institut fur Astrophysik, Karl-Schwarzschild-Str. 1, D-85748 Garching b. M\"unchen, Germany\\
}

\begin{document}

\maketitle

\label{firstpage}

\begin{abstract}
  We analyse spectroscopic measurements of 122 type Ia supernovae (SNe
  Ia) with $z<0.09$ discovered by the Palomar Transient Factory,
  focusing on the properties of the \Siii\ and \Caii\ `near-infrared
  triplet' absorptions. We examine the velocities of the photospheric
  \Siii, and the velocities and strengths of the photospheric and
  high-velocity \Caii, in the context of the stellar mass (\mstellar)
  and star-formation rate (SFR) of the SN host galaxies, as well as
  the position of the SN within its host. We find that SNe Ia with
  faster \Siii\ tend to explode in more massive galaxies, with the
  highest velocity events only occuring in galaxies with
  $\mstellar>3\times10^9$\,\msun.  We also find some evidence that
  these highest velocity SNe Ia explode in the inner regions of their
  host galaxies, similar to the study of \citet{2013Sci...340..170W},
  although the trend is not as significant in our data.  We show that
  these trends are consistent with some SN Ia spectral models, if the
  host galaxy stellar mass is interpreted as a proxy for host galaxy
  metallicity. We study the strength of the high-velocity component of
  the \Caii\ near-IR absorption, and show that SNe Ia with stronger
  high-velocity components relative to photospheric components are
  hosted by galaxies with low \mstellar, blue colour, and a high sSFR.
  Such SNe are therefore likely to arise from the youngest progenitor
  systems. This argues against a pure orientation effect being
  responsible for high-velocity features in SN Ia spectra and, when
  combined with other studies, is consistent with a scenario where
  high-velocity features are related to an interaction between the SN
  ejecta and circumstellar medium (CSM) local to the SN.
\end{abstract}


%%%%%%%%%%%%%%%%%%%%%%%%%%%%%%%%%%%%%%%%%%%%%%%%%%%%%%%%%%%%%%%%%%%%%%
\section{Introduction}
\label{sec:introduction}
%%%%%%%%%%%%%%%%%%%%%%%%%%%%%%%%%%%%%%%%%%%%%%%%%%%%%%%%%%%%%%%%%%%%%%
Type Ia supernovae (SNe Ia) are exceptional stellar explosions.  They
are believed to be the result of the thermonuclear explosion of an
accreting carbon-oxygen white dwarf star in a close binary system.
Although recent observations have constrained the size of the primary
star to be consistent with a compact object
\citep{2011Natur.480..344N,2012ApJ...744L..17B}, the nature of the
companion star that donates material is not yet clear. The various
possibilities include the classical single degenerate
\citep{1973ApJ...186.1007W} and double degenerate
\citep{1984ApJS...54..335I,1984ApJ...277..355W} scenarios, as well as
more contemporary variations on these themes. There is a varying
degree of evidence supporting both scenarios
\citep[e.g.,][]{2011Sci...333..856S,2012Sci...337..942D,2012Natur.481..164S};
for a recent review see \citet{2013arXiv1312.0628M}. A better
understanding of SN Ia progenitors would likely strengthen their
continuing use as cosmological probes
\citep[e.g.,][]{1998AJ....116.1009R,1999ApJ...517..565P,2009ApJS..185...32K,2007ApJ...659...98R,2011ApJ...737..102S,2013arXiv1310.3828R,2014arXiv1401.4064B}.

The host galaxies of SNe Ia are useful tools in these studies. For
example, studying the global properties of the host environments of
SNe Ia can reveal details of the progenitor systems and place broad
constraints on their ages
\citep[e.g.,][]{2005A&A...433..807M,2006ApJ...648..868S}. Previous
studies have also found significant correlations between SN Ia light
curve parameters and luminosities, and the properties of their hosts
\citep{1996AJ....112.2391H,2000AJ....120.1479H,2005ApJ...634..210G,2008ApJ...685..752G,2010ApJ...715..743K,2010ApJ...722..566L,2010MNRAS.406..782S,2011ApJ...743..172D,2012ApJ...755..125G,2013MNRAS.435.1680J,2013ApJ...764..191H,2013ApJ...770..108C,2013A&A...560A..66R,2014MNRAS.438.1391P}.
Intrinsically fainter SNe Ia (specifically those with faster light
curves) are preferentially located in massive/older galaxies than in
younger/lower-mass systems.  Galaxies with stronger star-formation
also tend to host slower, brighter SNe Ia than passive galaxies. From
a cosmological perspective, `corrected' SN Ia luminosities also show a
dependence on various parameters correlated with host galaxy stellar
mass (\mstellar) and metallicity, in the sense that brighter SNe Ia
tend to be found in massive/metal-rich galaxies. There are also
emerging trends that massive, or metal-rich, galaxies host redder SNe
Ia.

SN Ia spectral features are also important in understanding the
properties of the progenitor system, providing the only direct tracer
of the material in the SN ejecta. Work using the first large samples
of maximum-light SN Ia spectra
\citep[e.g.,][]{2005ApJ...623.1011B,2006PASP..118..560B} demonstrated
the existence of several sub-classes of SN Ia events.
\citet{2005ApJ...623.1011B} divided SNe Ia into three different groups
according to their spectral properties: A `FAINT' group including
SN\,1991bg-like events; a `high-velocity group' (HVG), which present a
high velocity gradient in their \Siii\ velocity evolution; and a
`low-velocity group' (LVG), which present a low velocity gradient in
their \Siii\ velocities.  \citet{2006PASP..118..560B} grouped SNe Ia
into four different groups according to the pseudo-equivalent-width
(pEW) of \Siiitmp\ and \Siii\ lines: A `core-normal' group, which
present homogeneous and intermediate pEWs; a `shallow-silicon' group,
which present low pEWs of both \Siii\ and \Siiitmp\ lines and include
SN\,1991T-like events (taken together, the core-normal and
shallow-silicon groups correspond to the
\citeauthor{2005ApJ...623.1011B} LVG); a `broad-line' group (similar
to the \citeauthor{2005ApJ...623.1011B} HVG), which present normal
pEWs of \Siiitmp\ but higher pEWs of \Siii; and a `cool' group
(similar to the \citeauthor{2005ApJ...623.1011B} FAINT group).

Such studies indicate that SNe Ia are not drawn from a one-parameter
family from the perspective of their spectral properties.
\citet{2007Sci...315..825M} studied the distribution of main elements
in nearby SNe Ia and found the outer \Siii\ velocities are similar
($\la12000$\,km\,s$^{-1}$) for all SNe Ia except those defined as
`HVG' group, which show much higher and dispersed \Siii\ velocities.
\citet{2009ApJ...699L.139W} divided SNe Ia into two groups according
to the photospheric velocities measured from \Siii\ absorptions. They
found that SNe Ia with high \Siii\ velocities (high-\vsiii; defined as
$\vsiii \ga 12,000$\,km\,s$^{-1}$) have a different extinction law
than normal-velocity events (normal-\vsiii; defined as $\vsiii <
12,000$\,km\,s$^{-1}$).  By applying different values of $R_V$ to each
group, the dispersion in the corrected SN peak luminosities can be
reduced. Many spectral indicators have also been found to correlate
with SN luminosity, and can be used for distance estimation
\citep{1995ApJ...455L.147N,2006MNRAS.370..299H,2009AAS...21332105B,2011A&A...526A..81B,2012MNRAS.425.1889S}.

However, there are fewer studies addressing the relationship between
SN Ia spectral features and their host galaxies. Early work
\citep{1993AJ....105.2231B} showed that, when considering the full
range of SNe Ia including subluminous events, SNe with the lowest Si
velocities tended to explode in early-type galaxies.
\citet{2012ApJ...748..127F} studied the relation between \Caii\ H\&K
velocity and host \mstellar, and found SNe Ia in massive galaxies have
lower \Caii\ H\&K velocities, although \citet{2012MNRAS.426.2359M}
suggested this could be caused by underlying relations between
light-curve width and \mstellar\ \citep[e.g.,][]{2010MNRAS.406..782S},
and between light-curve width and \Caii\ H\&K velocity
\citep[see][]{2012MNRAS.426.2359M,2014MNRAS.444.3258M}.
\citet{2013Sci...340..170W} found that high-\vsiii\ SNe Ia and
normal-\vsiii\ SNe Ia may originate from different populations with
respect to their radial distributions in their host galaxies.
High-\vsiii\ SNe Ia tend to concentrate in the inner regions of their
host galaxies, whereas the normal-\vsiii\ SNe Ia span a wider range of
radial distance.  Their result was interpreted as evidence for the
existence of two distinct populations of SNe Ia.

As well as photospheric spectral features, `high-velocity features'
(HVFs) in SN Ia spectra, particularly in the \Caii\ near-infrared
triplet, have also been studied
\citep{2003ApJ...591.1110W,2004ApJ...607..391G,2005ApJ...623L..37M,
  2005MNRAS.357..200M,2006ApJ...645..470T,2008ApJ...677..448T,2009A&A...508..229P,2013ApJ...770...29C,2013ApJ...777...40M,2014MNRAS.437..338C}.  The
physical origin of these HVFs is not yet clear, but it is generally
thought to be related to an abundance or density enhancement in the SN
ejecta, or interactions between the SN ejecta and a circumstellar
medium (CSM) local to the SN. Some interesting properties have been
found for these HVFs. For example, \citet{2014MNRAS.437..338C} show
the strength of the HVFs is connected to the decline rate of the SN
light curve: slower declining SNe have stronger HVFs. They also found
SNe Ia with stronger HVFs have lower \Siii\ photospheric velocities,
while the high-\vsiii\ SNe Ia discussed above show no distinct HVFs in
their maximum-light spectra.  HVFs could also provide different angles
to investigate the properties of CSM local to the SN.
\citet{2011Sci...333..856S} and \citet{2013MNRAS.436..222M} used the
narrow Na\,\textsc{i} D features as probes for CSM and found the SNe
presenting blue-shifted Na\,\textsc{i} D tend to be found in late-type
galaxies.  \citet{2012ApJ...752..101F} found these SNe with
blue-shifted Na\,\textsc{i} D generally have higher ejecta velocities
and redder colours at maximum light. Understanding the properties of
these HVFs could provide some clues to their relations with any CSM
and therefore the SN progenitor system.

In this paper, we use spectroscopic measurements of 122 low-redshift
SNe Ia discovered by the Palomar Transient Factory (PTF) to
investigate the relation between SN Ia spectroscopic properties and
the SN host galaxies. In particular, we focus on the properties of the
\Siii\ and \Caii\ near infrared absorptions. The spectral data and the
measurements themselves are presented in a companion paper
\citep{2014MNRAS.444.3258M}, while the host parameters were determined using
both photometric and spectroscopic data following
\citet{2014MNRAS.438.1391P}. In particular, we measure the SN--host
galaxy offset, the host galaxy stellar mass (\mstellar), the host
star-formation rate (SFR), the host gas-phase/stellar metallicities,
and the host mean stellar age.

A plan of the paper follows. In Section~\ref{sec:data} we introduce
the selection of our SN Ia spectral sample and the determination of
host parameters, and Section~\ref{sec:spectral-measurements} discusses
the spectral measurements. In Section~\ref{sec:results} we show the
results found between SN Ia spectral properties and host parameters.
The discussion and conclusion are presented in
Section~\ref{sec:discussion} and Section~\ref{sec:conclusions}.
Throughout this paper, we assume
$\mathrm{H_0}=70$\,km\,s$^{-1}$\,Mpc$^{-1}$ and a flat universe with
$\omatter=0.3$.

%%%%%%%%%%%%%%%%%%%%%%%%%%%%%%%%%%%%%%%%%%%%%%%%%%%%%%%%%%%%%%%%%%%%%%
\section{Data}
\label{sec:data}
%%%%%%%%%%%%%%%%%%%%%%%%%%%%%%%%%%%%%%%%%%%%%%%%%%%%%%%%%%%%%%%%%%%%%%

We begin by introducing the sample used in this work, including the
sample selection, SN light curve fitting and the determination of host
parameters.

%========================================
\subsection{The SN sample}
\label{sec:spectral-sample}
%========================================

The SNe Ia used in this work were discovered by the Palomar Transient
Factory (PTF), a project designed to explore the optical transient sky
using the CFH12k wide-field survey camera mounted on the Samuel Oschin
48-inch telescope (P48) at the Palomar Observatory
\citep{2008SPIE.7014E.163R,2009PASP..121.1334R,2009PASP..121.1395L}.
PTF searched in both $R$ and $g$-band filters (hereafter \rptf\ and
\gptf), and discovered $\sim1250$ spectroscopically confirmed SNe Ia
during its operation from 2009--2012.  SN candidates were identified
in image subtraction data and ranked using a machine learning
algorithm \citep{2012PASP..124.1175B}, and visually confirmed by
either members of the PTF collaboration or via the citizen science
project `Galaxy Zoo: Supernova' \citep{2011MNRAS.412.1309S}.

SN detections were then spectroscopically confirmed and followed-up
using a variety of facilities. These included: The William Herschel
Telescope (WHT) and the Intermediate dispersion Spectrograph and Image
System (ISIS), the Palomar Observatory Hale 200-in and the double
spectrograph \citep[DBSP;][]{1982PASP...94..586O}, the Keck-I
telescope and the Low Resolution Imaging Spectrometer
\citep[LRIS;][]{1995PASP..107..375O}, the Keck-II telescope and the
DEep Imaging Multi-Object Spectrograph
\citep[DEIMOS;][]{2003SPIE.4841.1657F}, the Gemini-N telescope and the
Gemini Multi-Object Spectrograph \citep[GMOS;][]{2004PASP..116..425H},
the Very Large Telescope and X-Shooter \citep{2011A&A...536A.105V},
the Lick Observatory 3m Shane telescope and the Kast Dual Channel
Spectrograph \citep{Kast_spectrograph}, the Kitt Peak National
Observatory 4m telescope and the Ritchey-Chretien Spectrograph, and
the University of Hawaii 88-in and the Supernova Integral Field
Spectrograph \citep[SNIFS;][]{2004SPIE.5249..146L}.  All of the
spectra used in this paper are available from the WISeREP archive
\citep{2012PASP..124..668Y}, and are presented in detail in
\citet{2014MNRAS.444.3258M}. Multi-colour light curves were not obtained by
default for all SNe; instead they were assembled in $g$, $r$ and $i$
via triggered observations on other robotic facilities, e.g. the
Liverpool Telescope \citep[LT;][]{2004SPIE.5489..679S}, the Palomar
60-in (P60) and the Las Cumbres Observatory Global Telescope Network
\citep[LCOGT;][]{2013arXiv1305.2437B} Faulkes telescopes (FTs).

We are interested in studying SNe Ia with well-observed optical
spectra near maximum light, and measuring their key spectral features.
Thus we make several selection cuts on the parent PTF sample. In
detail, these are as follows.

\begin{figure*}
	\centering
		\includegraphics*[scale=0.8]{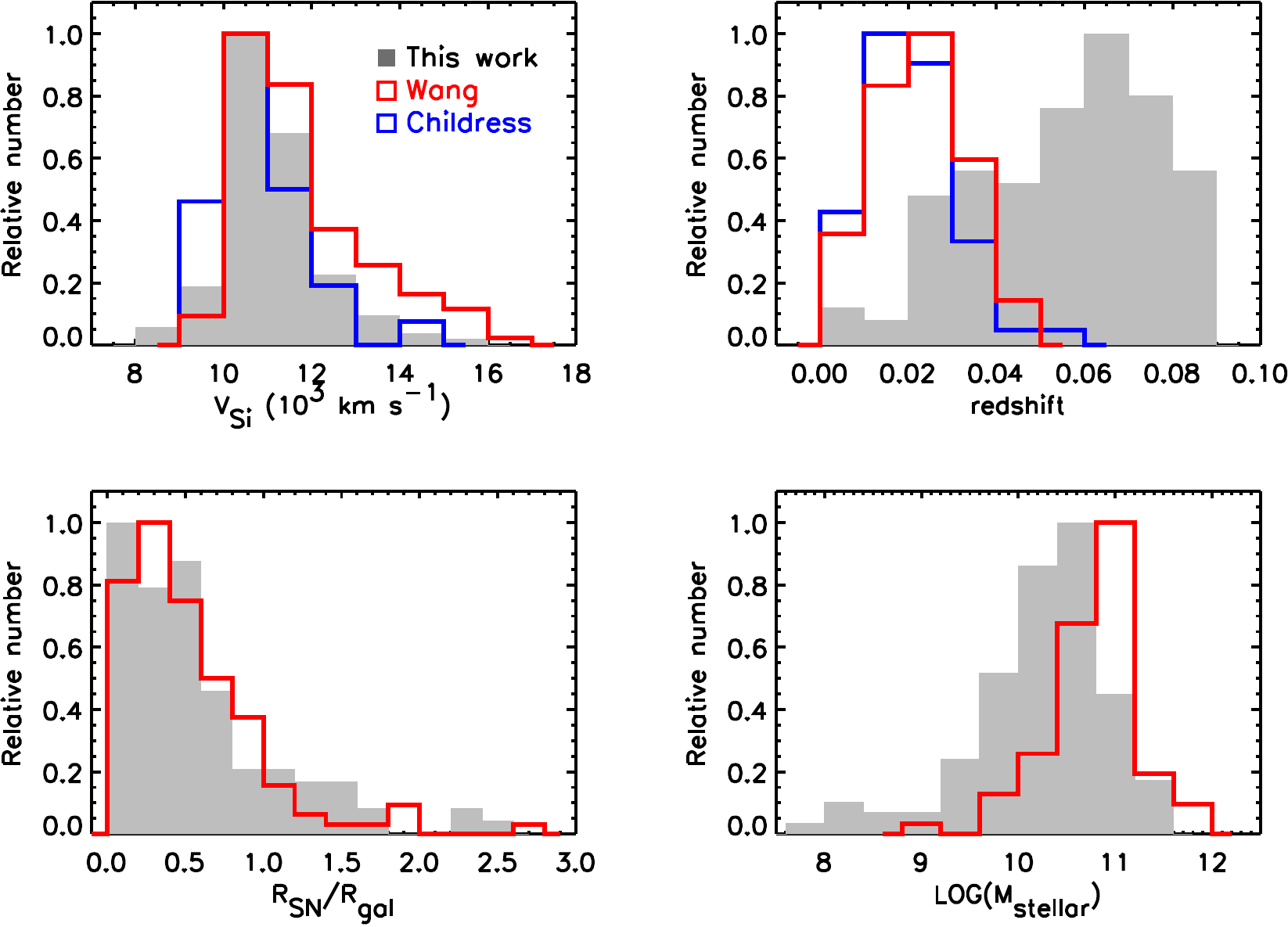}
                \caption{The distribution of the SN \Siii\ velocity,
                  redshift, host galaxy offset, and host stellar mass
                  (\mstellar) of our sample (filled histogram).  For
                  comparison, the sample used in
                  \citet{2013Sci...340..170W} and
                  \citet{2014MNRAS.437..338C} are also shown in red
                  and blue, respectively.}
        \label{sample_selection}
\end{figure*}

We first restrict our primary sample to those events with redshift of
$z<0.09$ (the final redshift distribution can be seen in
Fig.~\ref{sample_selection}).  This same redshift cut was made in
\citet{2014MNRAS.438.1391P}, and avoids most Malmquist-bias selection
effects (the median redshift of PTF SNe Ia is $\sim0.1$). It also
ensures that the \Caii\ near infrared triplet that we study here is
included in the spectral coverage (typically 3500 -- 9000\,\AA).
Secondly, we restricted the phases of SN spectroscopic observations to
be within 5 rest-frame days relative to $B$-band maximum light, where
the variation in the SN spectral velocities with phase show only a
relatively mild and linear trend (discussed further in
Section~\ref{sec:evolution-phase}).  We then restrict to the SNe where
the redshifts can be estimated from host galaxy features rather than
from SN template fitting, as spectral velocities are significantly
more uncertain in the latter case.  Finally, we exclude 21 SNe Ia from
our sample which have only a poor quality (low S/N) spectrum or light
curve.  In total, 122 events passed the above criteria and enter our
final sample.  We summarise the sample selection in
Table~\ref{selection}.

\begin{table}
\centering
\caption{The SN sample selection in this work.}
\begin{tabular}{lcc}
\hline\hline
 & Criterion & Num. of SNe left\\
\hline
PTF parent sample & -- & 1249 \\
Redshift cut & $\rm z<0.09$ & 527 \\
Spectral phase cut & $\rm -5\,d\leq t\leq5\,d$ & 160 \\
Redshift derived from host & -- & 143 \\
LC quality cut & -- & 133 \\
Spectrum quality cut & -- & 122 \\
(In SDSS) & -- & 100 \\
\hline
\end{tabular}
\label{selection}
\end{table}

Of the 122 SNe Ia, 100 have host galaxies and available $ugriz$
photometry in the Sloan Digital Sky Survey (SDSS) Data Release 10
\citep[DR10;][]{2013arXiv1307.7735A}, which we use for an \mstellar\
determination (Section~\ref{sec:host}). A further two events lie
outside of the SDSS footprint but their \mstellar\ can be determined
using LT images taken as part of the SN photometric follow-up
campaign. The details of calibrating these LT photometry can be found
in \citet{2014MNRAS.438.1391P}. The remainder of the 20 SNe without multi-colour
photometric host data lie outside of the SDSS footprint.

The SiFTO light curve fitting code \citep{2008ApJ...681..482C} was
used to fit the SN light curves.  The SN stretch ($s$), $B-V$ colour
at $B$-band maximum light (\col), the rest-frame $B$-band apparent
magnitude at maximum light ($m_B$), and the time of the maximum light
in the rest-frame $B$-band are determined.  Further details about the
SN light curve fitting can be found in \citet{2014MNRAS.438.1391P}.

%========================================
\subsection{Host galaxy properties}
\label{sec:host}
%========================================

The main aim of this work is to investigate the relations between SN
Ia spectral properties and their host parameters. The host stellar
mass (\mstellar) and SN offset were determined using the photometric
data, and some of the hosts in our sample also have spectral
parameters as measured in \citet{2014MNRAS.438.1391P}.  The detailed
determination of host parameters were described in
\citet{2014MNRAS.438.1391P}. We summarise briefly as follows.

We determined the host \mstellar\ and star-formation rate (SFR) using
the photometric redshift code \textsc{z-peg}
\citep{2002A&A...386..446L}.  \textsc{z-peg} fits the observed colour
of the galaxies with galaxy Spectral Energy Distributions (SEDs) from 9
different spectral types (SB, Im, Sd, Sc, Sbc, Sb, Sa, S0 and E). Milky
Way extinction is corrected for, and a further foreground dust screen
varying from $E(B-V)=0$ to 0.2\,mag in steps of 0.02\,mag is fit.
Throughout the paper a \citet{1955ApJ...121..161S} Initial Mass
Function (IMF) is assumed.

Of the 122 SN Ia host galaxies studied in this work, 41 events were
studied in \citet{2014MNRAS.438.1391P}, and therefore
have well-measured spectral SFRs, gas-phase/stellar
metallicities, and stellar ages.  For these objects, the codes
\textsc{ppxf} \citep{2004PASP..116..138C} and \textsc{gandalf}
\citep{2006MNRAS.366.1151S} were used to fit the host spectrum based
on the stellar templates provided by the \textsc{miles} empirical
stellar library \citep{2006MNRAS.371..703S,2010MNRAS.404.1639V}. The
potential AGN hosts in our sample were identified using the diagnostic studied
by \citet*{1981PASP...93....5B} (the so-called BPT diagram) with the
criterion proposed by \citet{2001ApJ...556..121K}, and are not used
for further emission-line analyses. The SFR is determined from the
H${\alpha}$ luminosity based on the conversion of
\citet{1998ARA&A..36..189K}.  Following the procedure described in
\citet{2008ApJ...681.1183K}, we adopt the metallicity calibration studied by
\citep[][hereafter PP04]{2004MNRAS.348L..59P} to calibrate the gas-phase
metallicity. The `N2' method (using the line ratio [\nii] $\lambda 6584$/H$\alpha$)
in PP04 calibration is used.
The mass-weighted stellar metallicity and age are
determined using the `full spectrum fitting' method, and \textsc{ppxf}
is used to fit the stellar continuum of our host spectra.  The stellar
metallicity and age were then estimated by using a weighted average of
the model templates.

We also measure the SN offset (\rsn) from its host galaxy.  \rsn\ is
defined as the separation (i.e., projected radial distance) between
the SN position and the host galaxy centre.  The coordinates of the SN
position were measured as a product of the P48 SN photometry
procedures, and the host centre was determined using
\textsc{sextractor} \citep{1996A&AS..117..393B} on the \rptf\ or
\gptf\ reference images.  We examine the potential hosts in the
reference images by cross-checking the same field in the SDSS
database; therefore we only measured the hosts which are covered by
the SDSS footprint.  For two SNe in our sample, we were not able to
constrain the host positions as they were either too close to a bright
star or in a very crowded field.  This gave \rsn\ measures for 98
SNe.

\begin{figure*}
	\centering
		\includegraphics*[scale=0.85]{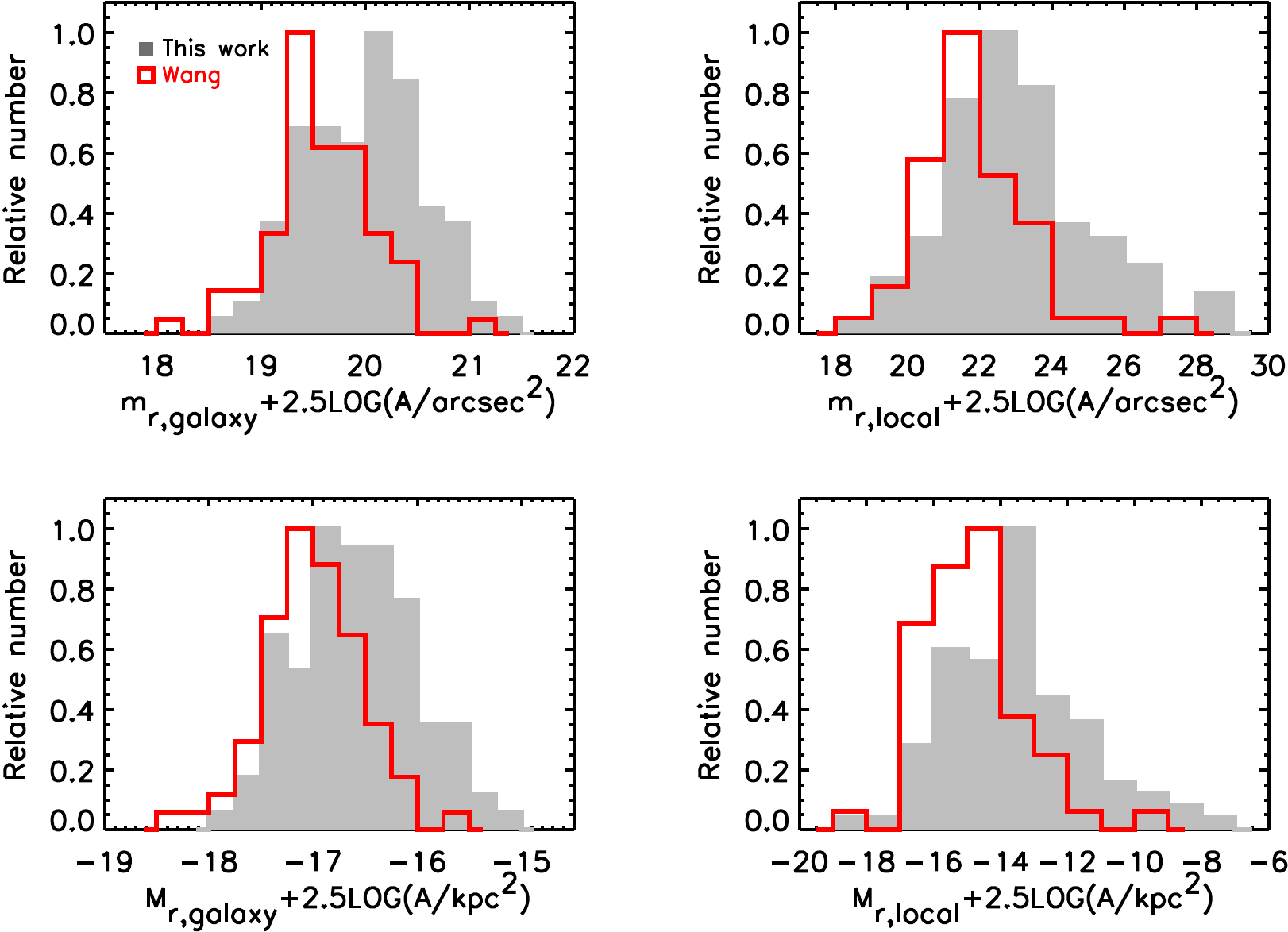}
                \caption{The comparison of host galaxy surface
                  brightness from the SNe Ia in this work and that of
                  \citet{2013Sci...340..170W}. Top panels: The
                  distribution of surface brightness considering the
                  whole galaxy (left panel) and the local region (a
                  circle of radius $1\arcsec$) at the SN position
                  (right panel), respectively.  The surface brightness
                  is computed as $m_r+2.5\log A$, where $A$ defines an
                  area which encloses either the whole galaxy or the
                  local region at the SN position.  The enclosed area
                  is in units of arcsec$^2$. Bottom panels: The same
                  as the top panels, but considering the absolute
                  magnitude ($M_r$) with enclosed area in units of
                  kpc$^2$.}
        \label{SB_compare}
\end{figure*}

These offsets are projected offsets, and so we investigate the
possibility of deprojecting the offsets using our data. We can
approximately deproject the SN offset using the position angle and
axial ratio of the host galaxy measured by \textsc{sextractor}
\citep[following the procedure described in][]{2009A&A...508.1259H}.
This correction is only valid for disk galaxies with moderate
inclinations, and no corrections should be made for elliptical
galaxies or disk galaxies with very large inclinations (i.e., nearly
edge-on).  Compared to the $z<0.05$ host galaxies studied in
\citet{2013Sci...340..170W}, our host galaxies are generally more
distant (Fig.~\ref{sample_selection}) and smaller in their apparent
sizes.  This introduces difficulties in classifying these galaxies and
thus further increases the uncertainties in deprojecting the SN
offsets.  Furthermore, we found our results presented in later
sections are not sensitive to the correction, and therefore no
deprojection is applied for the host galaxies in this work.

We also measure a typical size of each of the host galaxies \rgal,
defined as the radius at which 90\% of the flux from the galaxy is
enclosed. We then use \rgal\ to normalise the SN offset. This is
similar to the approach of \citet{2013Sci...340..170W} but differs in
detail; they use the $B=25$\,mag\,arcsec$^{-2}$ isophote as \rgal. As
this definition may introduce a redshift-dependent bias in the sizes
measured, we do not attempt to replicate this measure on our
\rptf\ and \gptf\ images.

Finally, we determined the surface brightness of our host galaxies and
those from \citet{2013Sci...340..170W} using SDSS $r$-band images.
The galaxy surface brightness is computed as $m_r+2.5\log A$, where
$A$ represents the area of the ellipse which encloses the whole galaxy
(in units of arcsec$^2$), and $m_r$ is the apparent magnitude of the
host galaxy determined by integrating the total counts enclosed within
that ellipse (FLUX\_AUTO in \textsc{sextractor}).  We also compute a
`local' surface brightness by integrating the total counts within a
circular aperture of radius $1\arcsec$ centred at the SN position.  We
then express the surface brightness in units of absolute magnitude
($M_r$) per kpc$^2$. The results can be found in
Fig.~\ref{SB_compare}.

%%%%%%%%%%%%%%%%%%%%%%%%%%%%%%%%%%%%%%%%%%%%%%%%%%%%%%%%%%%%%%%%%%%%%%
\section{SPECTRAL MEASUREMENTS}
\label{sec:spectral-measurements}
%%%%%%%%%%%%%%%%%%%%%%%%%%%%%%%%%%%%%%%%%%%%%%%%%%%%%%%%%%%%%%%%%%%%%%

In this section, we discuss our method to measure the SN spectral
features. The key features of interest are the pseudo-equivalent
widths (pEW) and velocities of \Siii\ and the \Caii\ near infrared
(NIR) triplet.

%================================
\subsection{Line measurement}
\label{sec:line-measure}
%================================

A complete description of the spectral feature measurement techniques
can be found in \citet{2014MNRAS.444.3258M}. The method is similar to that of
\citet{2014MNRAS.437..338C}, but with some differences, and developed
independently. Here we briefly summarise our procedure.

We started by fitting the \Siii\ doublet, a prominent line in SN Ia
spectra with little contamination from other features. We correct the
SN spectrum into the rest frame, define (by hand) continuum regions on
either side of the feature, and fit a straight line pseudo-continuum
across the absorption feature. The feature is then normalised by
dividing it by the pseudo-continuum. A double-Gaussian fit is
performed to the normalised \Siii\ doublet line in velocity space
using the \textsc{mpfit} package \citep{2009ASPC..411..251M}. The
centres of the two Gaussians have a fixed velocity difference
corresponding to the difference in the wavelengths of the doublet
($\lambda$6347\AA\ and $\lambda$6371\AA) and the same width.
The relative strengths of the individual line components are fixed
to be equal by assuming an optically thick regime 
\citep[e.g.,][]{2014MNRAS.437..338C,2014MNRAS.444.3258M}.
The resulting fit then gives the velocity of the feature, together 
with the pEW.

We then fit the complex \Caii\ NIR feature. This is a more complicated
task as the \Caii\ high-velocity component can be strong around SN
maximum light, and thus we fit for both photospheric velocity and
high-velocity components in the absorption complex.  To achieve this,
the \Siii\ velocity and width derived from previous fit are used as
initial guesses for the photospheric component of the \Caii\ NIR line.  We
then require the velocity of the \Caii\ NIR photospheric component to
be within 25\% of the \Siii\ velocity, and the \Caii\ NIR
high-velocity component to be larger than the \Siii\ velocity by at
least 2000\,$\mathrm{km\,s^{-1}}$.  No other constraints are applied
during the fit.

To ensure our measurements of these absorption lines are not sensitive
to the locations we selected for deriving the pseudo-continuum, we
randomly move the red and blue pseudo-continuum regions by up to
10\AA\ with respect to the original location, and re-fit the line
profile. This random process is repeated 200 times.  The final values
reported in this work are the mean of the measurements from all the
iterations, and the uncertainty is the standard deviation.

%================================
\subsection{Phase evolution of spectral features}
\label{sec:evolution-phase}
%================================

\begin{table}
\centering
\caption{The gradient of the SN spectral feature evolution in this work.}
\begin{tabular}{lc}
\hline\hline
Feature & Gradient\\
\hline
\Siii\ velocity & $-32\pm22$ km\,$\mathrm{s^{-1}d^{-1}}$ \\
\Caii\ NIR velocity (PVF) & $+38\pm59$ km\,$\mathrm{s^{-1}d^{-1}}$ \\
\Caii\ NIR velocity (HVF) & $-336\pm80$ km\,$\mathrm{s^{-1}d^{-1}}$ \\
\Caii\ NIR pEW (PVF) & $+3.2\pm1.1$\,\AA\ $\mathrm{d^{-1}}$ \\
\Caii\ NIR pEW (HVF) & $-3.5\pm2.1$\,\AA\ $\mathrm{d^{-1}}$ \\
\hline
\end{tabular}
\label{gradient}
\end{table}

\begin{figure}
  \centering
  \includegraphics*[scale=0.47]{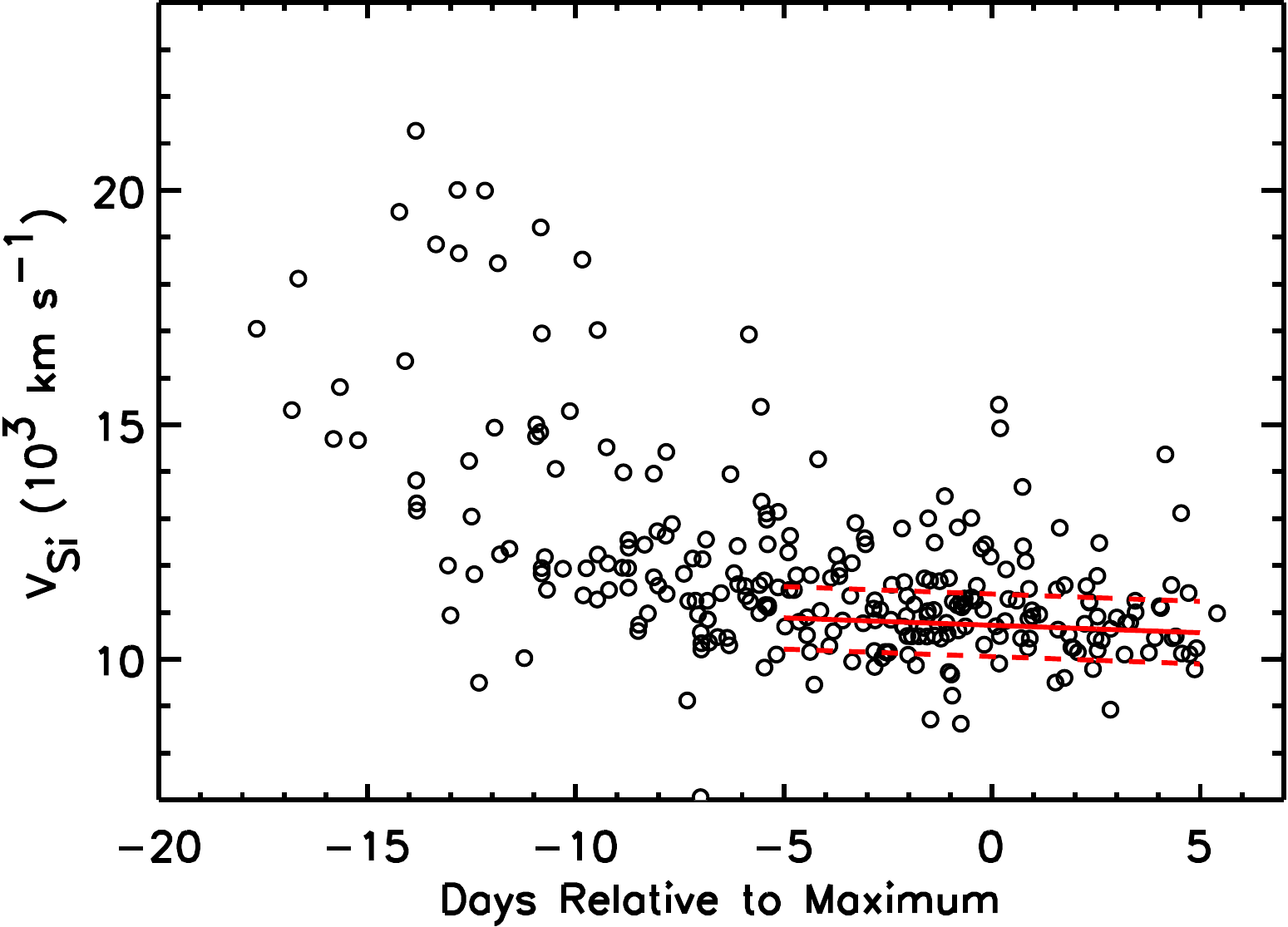}
  \caption{The \Siii\ velocity as a function of the phase of the SN Ia
    spectroscopic observation.  The red solid line shows a linear fit
    to data with phases from $-5$\,d to $+5$\,d. The dashed lines show
    the 1-$\sigma$ range relative to the fit.}
  \label{phase-vsi6150}
\end{figure}

As discussed in Section~\ref{sec:spectral-sample}, we restrict the
phases of our PTF SN Ia spectral sample to be within 5 days of the
$B$-band maximum light of the SN. Over this phase range, SN Ia
spectral velocities generally show a mild and linear trend
\citep[e.g.,][]{2012MNRAS.425.1819S}.  This can also be seen in our
sample in Fig.~\ref{phase-vsi6150} \citep[see also][]{2014MNRAS.444.3258M},
which shows the \Siii\ velocity as a function of phase.
We list the gradient of each spectral feature in this work in Table~\ref{gradient}.
However, the rate of phase evolution for these spectral features could
show considerable diversity for different SNe Ia 
\citep{2005ApJ...623.1011B,2012AJ....143..126B}.
Therefore we did not correct the SN spectral features using the gradients 
listed in Table~\ref{gradient}, and instead chose to use the spectral 
measurements with the phases closest
to the time of $B$-band maximum light.
We found our results in later
sections are not sensitive to the phase corrections.

%========================================
\subsection{Comparison with other samples}
\label{sec:compare-other-sample}
%========================================

Fig.~\ref{sample_selection} and Fig.~\ref{SB_compare} show the
comparison of the \Siii\ velocity (\vsiii), redshift, normalised SN
offset (\offset), \mstellar\ and surface brightness distributions
between this work and the sample used in \citet{2013Sci...340..170W}
and \citet{2014MNRAS.437..338C}. Here, we only show the 123
`Branch-Normal' SNe Ia used in \citet{2013Sci...340..170W}. We
determined the \mstellar\ for the hosts studied in
\citet{2013Sci...340..170W} using the same method described in
Section~\ref{sec:host}, with 74 out of 123 events having available
SDSS photometry for comparison to our sample.

The \citet{2013Sci...340..170W} \mstellar\ distribution is very
different from this work.  This is almost certainly due to the
selection of the SN sample used by \citet{2013Sci...340..170W}, which
were discovered by the Lick Observatory Supernova Search (LOSS),
designed as a galaxy-targeted survey. A Kolmogorov-Smirnov (K-S) test
gives a $<1$ per cent probability that the \mstellar\ distributions of
\citet{2013Sci...340..170W} and this work are drawn from the same
underlying population.

We also see a large difference in the \vsiii\ distributions, with
\citet{2013Sci...340..170W} having a larger fraction of SNe Ia with
high \vsiii. Our \vsiii\ distribution is consistent with that of
\citet{2014MNRAS.437..338C}. We find no evidence for a redshift
evolution in \vsiii\ in our sample, making the small redshift
difference between our sample and that of \citet{2013Sci...340..170W}
unlikely to drive the offset in Fig.~\ref{sample_selection}. We
discuss the possible origin of the \vsiii\ discrepancy in
Section~\ref{sec:offset}.  The SN radial distribution of this work is
consistent with that of \citet{2013Sci...340..170W}, despite the
slightly different definitions (Section~\ref{sec:host}).

Our sample also differs from \citet{2013Sci...340..170W} in the
surface brightness distributions of the host galaxies
(Fig.~\ref{SB_compare}).  The majority of the host galaxies sampled by
\citet{2013Sci...340..170W} have high surface brightnesses, where in
this work we sample galaxies with both high and low surface
brightness.  A K-S test gives a $<0.01\%$ probability that the host
galaxies studied from \citet{2013Sci...340..170W} and this work are
drawn from the same population with respect to their surface
brightness.  Similar results were also found when investigating the
local surface brightness at the SN position, with the sample in
\citet{2013Sci...340..170W} again biased toward those with high
surface brightness. However, we find no evidence that PTF is biased
against events on a high-surface brightness background; the bright
ends of the surface brightness distributions are similar between PTF
and \citet{2013Sci...340..170W}.

\begin{figure*}
	\centering
		\includegraphics*[scale=0.74]{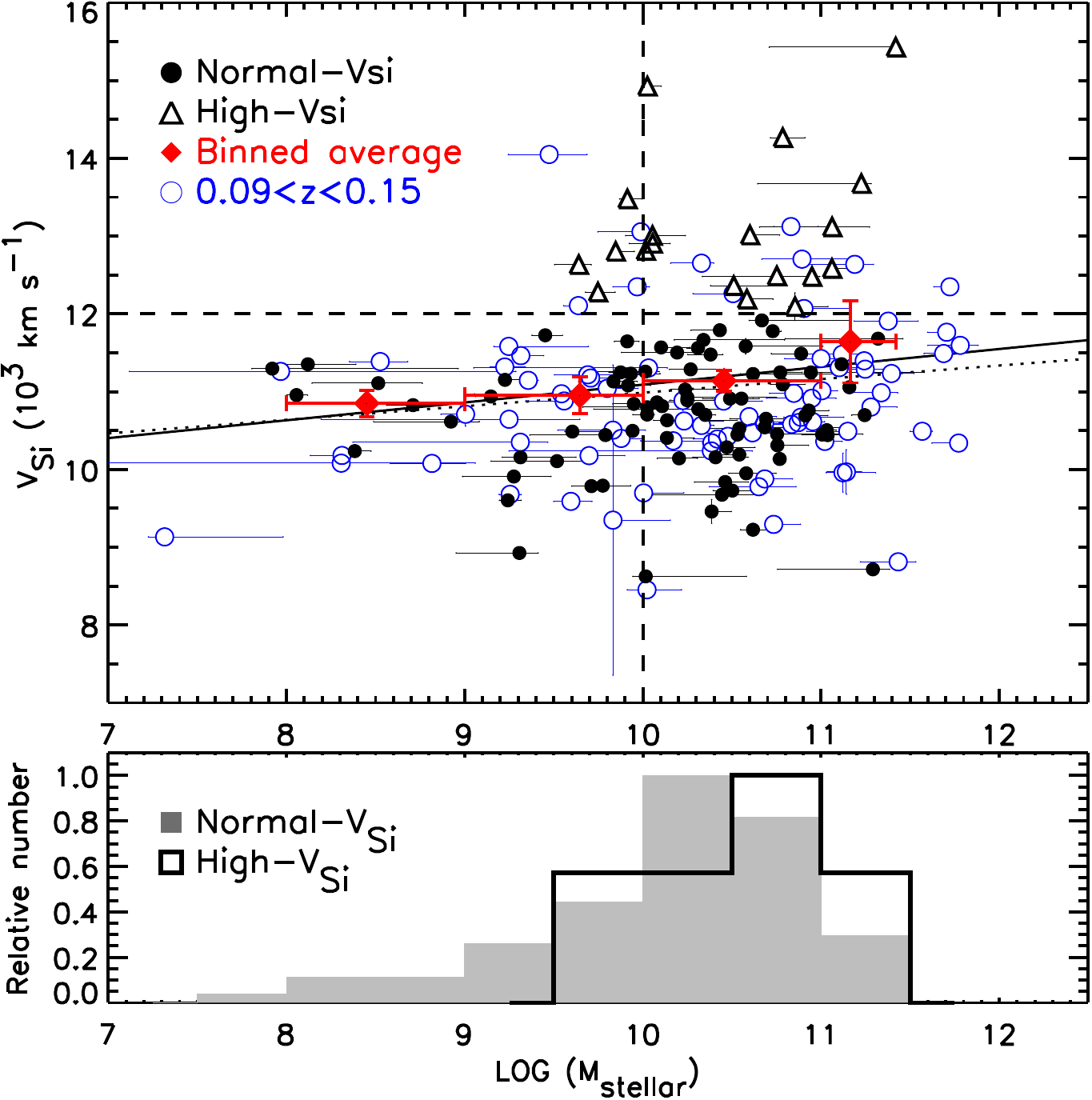}
                \caption{The \Siii\ velocities (\vsiii) as a function
                  of \mstellar.  The high-\vsiii\ SNe Ia (with
                  $\vsiii\geq12000\,\mathrm{km\,s^{-1}}$) are shown as
                  open triangles, and the normal-\vsiii\ SNe Ia are
                  shown as filled circles. The red diamonds represent
                  the mean velocities in bins of \mstellar, and their
                  error bars are the width of the bins and the error
                  on the mean. The vertical and horizontal dashed
                  lines represent the criterion used to split the
                  sample in velocity and \mstellar\ space,
                  respectively. The solid line is the linear fit to
                  the data in the plot (filled circles plus open
                  triangles).  We overplot the SNe with $0.09<z<0.15$
                  in open blue circles for comparison.  The linear fit
                  to all the data (including those $0.09<z<0.15$) is
                  shown in dotted line.  The bottom histograms show
                  the \mstellar\ distributions of high-\vsiii\ and
                  normal-\vsiii\ SNe Ia.  }
        \label{m-vsi}
\end{figure*}

\begin{figure*}
	\centering
		\begin{tabular}{c}
		\includegraphics*[scale=0.5]{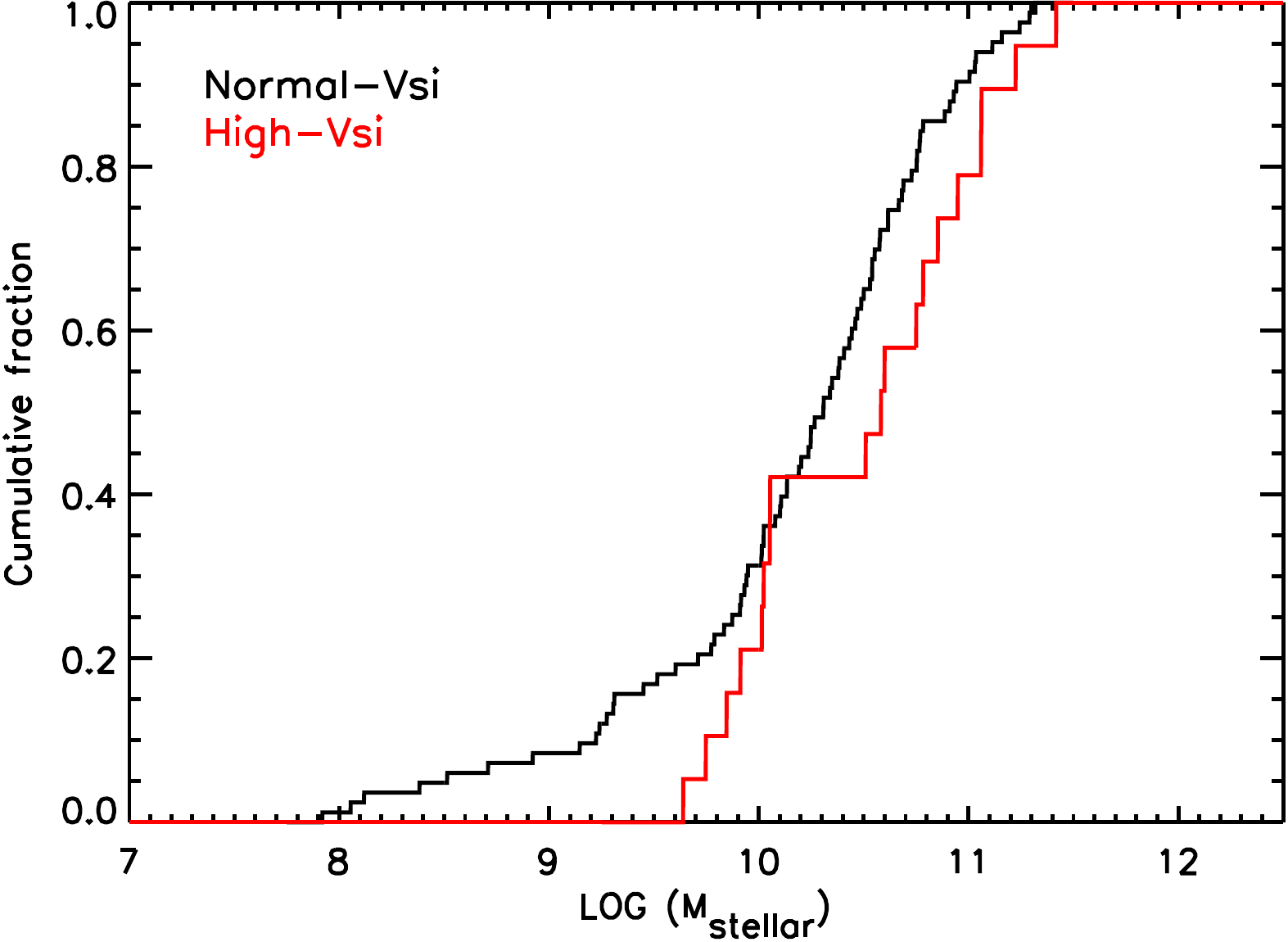}
		\includegraphics*[scale=0.5]{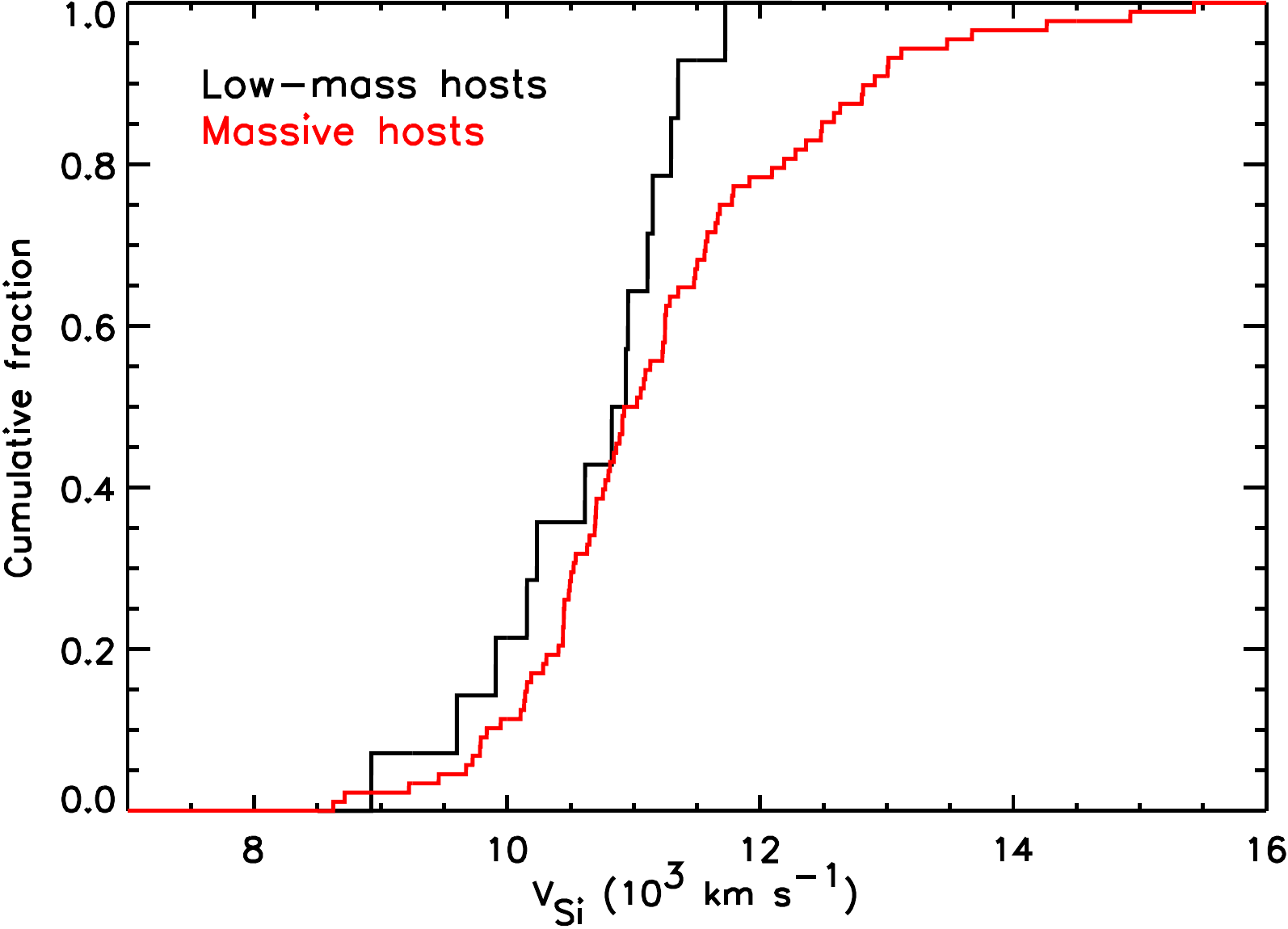}
		\end{tabular}
		\caption{Left panel: The cumulative fractions of host
                  \mstellar\ of high-\vsiii\ and normal-\vsiii\ SNe
                  Ia. Right panel: The cumulative fractions of \vsiii\
                  of SNe Ia in high- and low-mass host galaxies.}
		\label{cdf1}
\end{figure*}

%%%%%%%%%%%%%%%%%%%%%%%%%%%%%%%%%%%%%%%%%%%%%%%%%%%%%%%%%%%%%%%%%%%%%%
\section{RESULTS}
\label{sec:results}
%%%%%%%%%%%%%%%%%%%%%%%%%%%%%%%%%%%%%%%%%%%%%%%%%%%%%%%%%%%%%%%%%%%%%%

We now turn to the main results of our study. We split our analysis
into two parts. The first concerns the properties of the photospheric
features and velocities, namely the \Siii\ and \Caii\ NIR photospheric
velocities (\vsiii\ and \vcaii). In this discussion we include trends
for SNe with very high \vsiii, which we denote high velocity events
and, following \citet{2013Sci...340..170W}, define as
$\vsiii\geq12000\,\mathrm{km\,s^{-1}}$. The second part concerns the
strength of the high-velocity features present in the \Caii\ NIR
absorption.

%=================================
\subsection{Photospheric features and host galaxy parameters}
\label{sec:pvfs-host}
%=================================

%------------------------
\subsubsection{Stellar mass}
\label{sec:mass}
%------------------------

We begin by examining trends with host galaxy stellar mass
(\mstellar). Fig.~\ref{m-vsi} shows \vsiii\ as a function of
\mstellar.  There is a broad trend that high-\vsiii\ SNe Ia tend to
reside in massive galaxies, whereas the normal-\vsiii\ SNe Ia
($\vsiii<12000\,\mathrm{km\,s^{-1}}$) are found in galaxies of all
mass.  In our sample, high-\vsiii\ SNe Ia only occur in galaxies with
$\log(\mstellar)>9.5$. Fig.~\ref{cdf1} (left panel) shows the
cumulative distribution function of host galaxy \mstellar\ for both
high- and normal-\vsiii\ SNe Ia. The two distributions are different
particularly at the low-\mstellar\ end.  The cumulative distribution
function of \vsiii\ for SNe Ia in high-\mstellar\ ($\log(M/M_{\odot}) >10$) 
and low-\mstellar\ ($\log(M/M_{\odot})<10$) hosts is
also shown in Fig.~\ref{cdf1} (right panel).  The distribution of
high-\mstellar\ hosts is different from low-\mstellar\ hosts due to
additional SNe at high \vsiii.

The classical K-S test is less sensitive when testing two
distributions that vary mainly in their tails, and indeed gives a
$p$-value of 0.25 that the normal- and high-\vsiii\ SN \mstellar\
distributions are drawn from the same underlying population
(i.e., they are not two distinct populations).  
The same test gives a $p$-value of 0.21 that
low-\mstellar\ and high-\mstellar\ hosts have \vsiii\ distributions
drawn from the same population. Drawing 19 SNe Ia, the size of the
high-\vsiii\ sample, with replacement and at random from the full PTF
sample, only in 3.1\% of iterations do all the selected SNe have
\mstellar\ greater than $4.36\times10^9\,\msun$ (the minimum
\mstellar\ of the high-\vsiii\ SNe Ia in this work). Finally, a
linear fitting performed by \textsc{linmix}
\citep{2007ApJ...665.1489K} shows a non-negative slope at 87\%
probability.

These results are suggestive, but not conclusive, that high-\vsiii\
SNe Ia prefer high-\mstellar\ host galaxies, and of a more general
relationship between \vsiii\ and \mstellar. We investigate if the
trend holds when adding additional high-redshift $0.09\leq z<0.15$ PTF
SNe Ia (our primary sample has $z<0.09$; Section
\ref{sec:spectral-sample}). This larger sample has more significant
trends: the linear fitting prefers a non-negative slope at 93\%
probability.

Previous studies found some evidence that \vcaiihk\ correlates with
\mstellar, in the sense that SNe Ia in more massive galaxies tend to
have lower \vcaiihk\
\citep[e.g.,][]{2012ApJ...748..127F,2012MNRAS.426.2359M}; note this is
opposite to the trends with \vsiii\ in Fig.~\ref{m-vsi}).
\citet{2012MNRAS.426.2359M} suggested that this trend is caused by an
underlying relationship between SN light curve shape and \Caii\ H\&K
velocity, in the sense that SNe Ia with higher stretches tend to have
higher \Caii\ H\&K velocities. They found the trend between \Caii\
H\&K velocity and \mstellar\ disappeared after removing the
correlation between SN light curve shape and \Caii\ H\&K velocity.

However, \citet{2013MNRAS.435..273F} argued the contribution of
\Siiihk\ line in \Caii\ H\&K feature is important, and may make the
measurement of the \Caii\ H\&K line uncertain.  They suggested the blue
component of the \Caii\ H\&K feature is caused by the \Siiihk\ feature
for most SNe Ia, and that therefore the correlation between SN light
curve shape and \vcaiihk\ observed in \citet{2012MNRAS.426.2359M}
could be due to the strong correlation between \Siiihk\ and excitation
temperature, and therefore the SN light curve shape.
\citet{2014MNRAS.437..338C} found the HVFs are stronger in higher
stretch SNe Ia, supporting the idea proposed by
\citet{2012MNRAS.426.2359M}. However, they also showed the \Siiihk\
line does have some impact on the \Caii\ H\&K line.  Taken together,
these results imply the trend between \vcaiihk\ and \mstellar\ may not
be unambiguously caused by an intrinsic property of \Caii\ feature
itself.

\begin{figure}
	\centering
		\includegraphics*[scale=0.64]{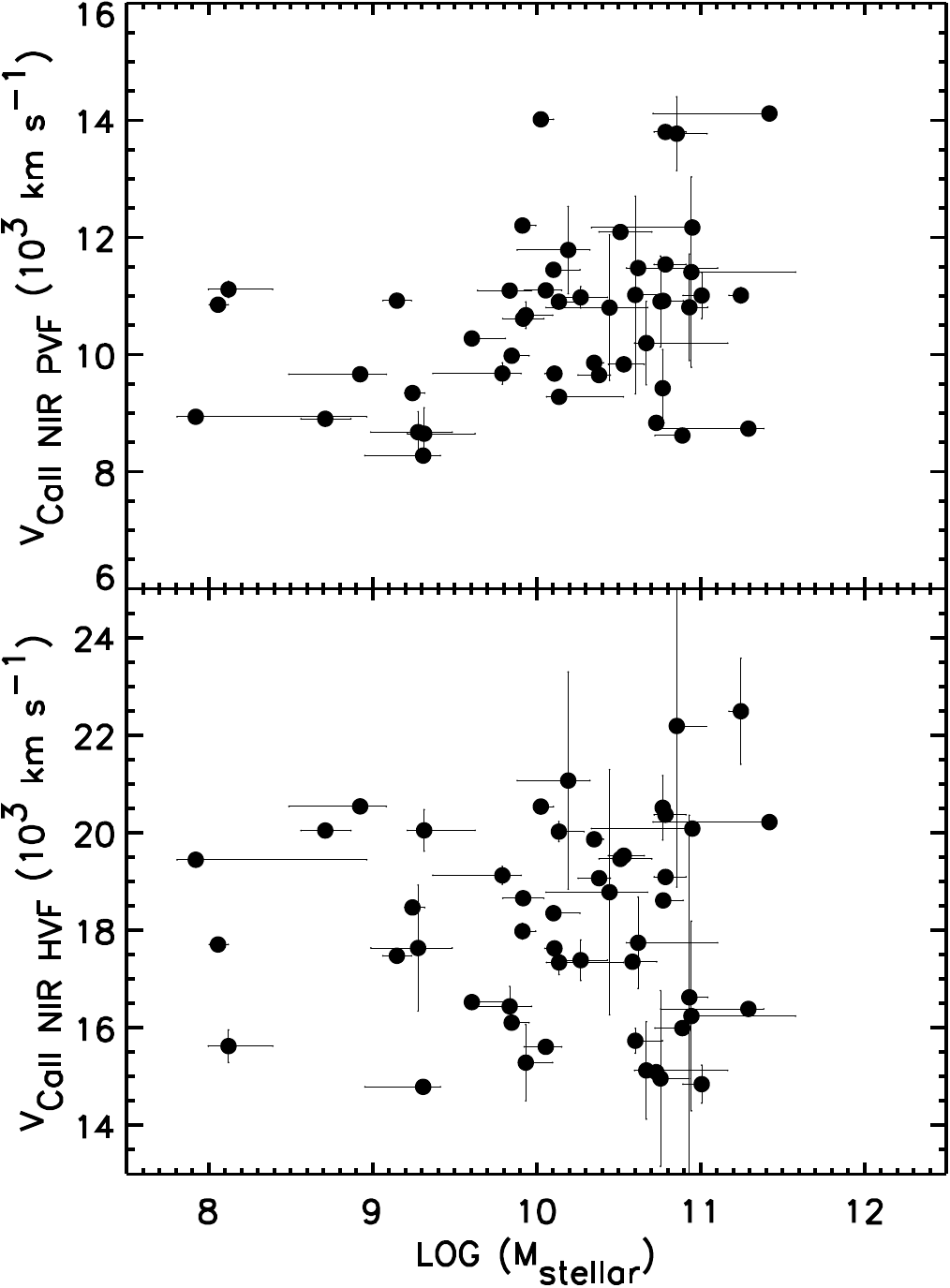}
                \caption{The upper panel shows the \Caii\ NIR
                  velocities of photospheric feature (PVF) component
                  as a function of \mstellar. The lower panel shows
                  the \Caii\ NIR velocities of high-velocity feature
                  (HVF) component as a function of \mstellar.  }
        \label{m-vcair}
\end{figure}

Clearly a problem with these analyses is the use of the \Caii\ H\&K
feature, which is a difficult feature to model accurately. Here we use
\Caii\ NIR line instead, as it provides a cleaner measurement of
\Caii\ velocity without contamination from other features (e.g., Si);
we are also able to more easily decompose the high-velocity and
photospheric-velocity components.  Fig.~\ref{m-vcair} shows the \Caii\
NIR velocity of these photospheric and high-velocity components as a
function of \mstellar.  Similar to our \vsiii\ analysis, SNe with the
highest photospheric \vcaii\ also tend to reside in massive galaxies.
The high-velocity \vcaii\ shows no significant trend with \mstellar,
although there is a suggestion that SNe with low \vcaii\ are deficient
in lower-mass galaxies.  In our companion paper, \citet{2014MNRAS.444.3258M},
we show that SNe Ia with strong HVFs in the \Caii\ NIR feature
relative to photospheric \Caii\ generally have higher \Caii\
velocities.  If the hosts with lower \mstellar\ ($\rm \log M < 10$)
are dominated by SNe with strong HVFs (as we will see in
Section~\ref{sec:hvfs-host}), we will expect the SNe in low \mstellar\
galaxies to have higher \Caii\ NIR HVF velocities.  This may also
explain the tendency for the low-mass galaxies in our sample to lack
SNe with low \Caii\ NIR HVF velocities.

\begin{figure*}
	\centering
		\includegraphics*[scale=0.74]{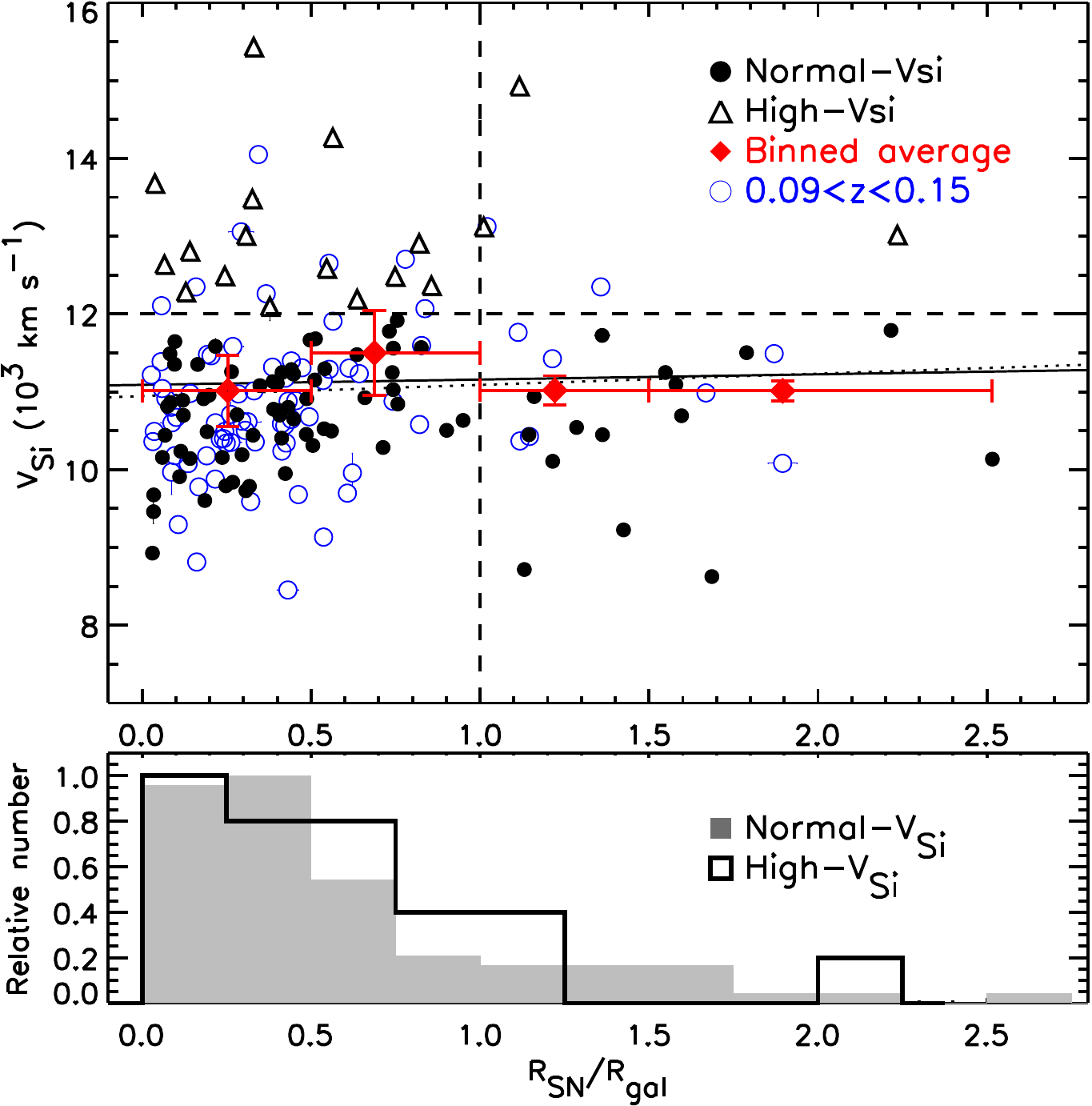}
                \caption{As Fig.~\ref{m-vsi}, but considering
                the \offset\ ratio instead of \mstellar.
                R${_\mathrm{SN}}$ and R${_\mathrm{gal}}$ are defined in Section~\ref{sec:host}.}
        \label{offset-vsi}
\end{figure*}

\begin{figure*}
	\centering
		\begin{tabular}{c}
		\includegraphics*[scale=0.5]{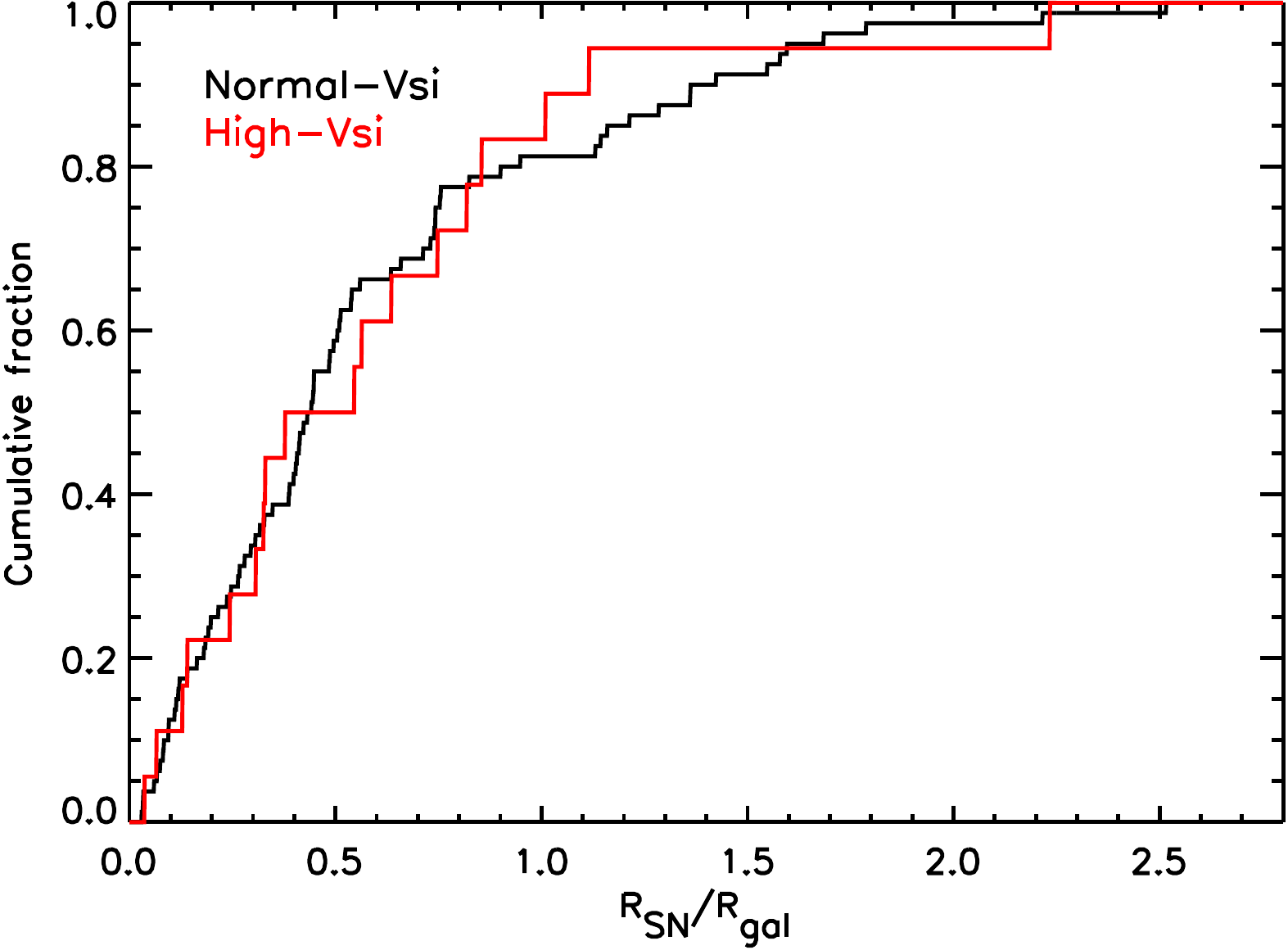}
		\includegraphics*[scale=0.5]{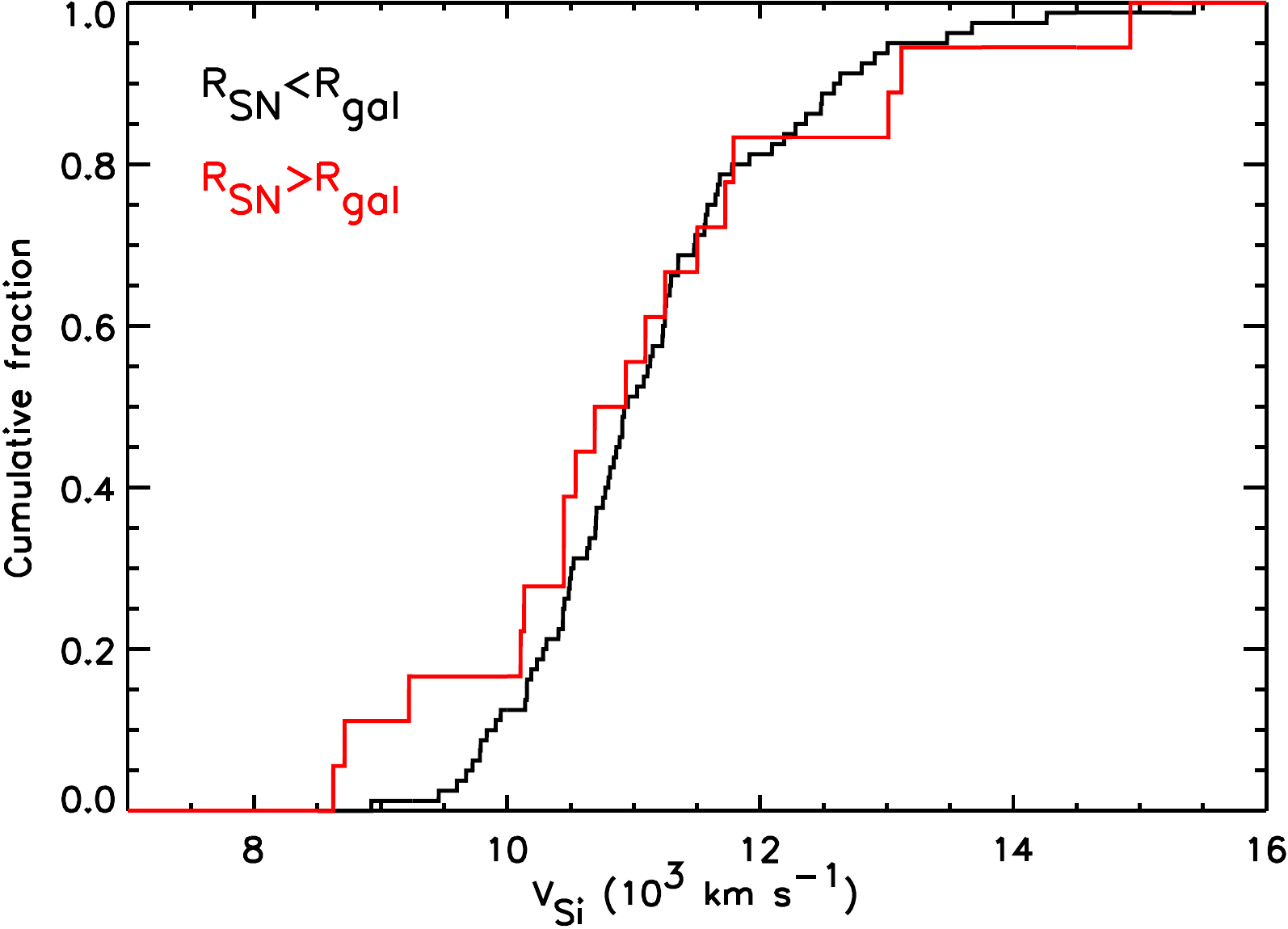}
		\end{tabular}
		\caption{Left panel: The cumulative fractions of
                  \offset\ of high-\vsiii\ and normal-\vsiii\ SNe
                  Ia. Right panel: The cumulative fractions of in
                  \vsiii\ of SNe Ia in larger radial distance and
                  smaller radial distance.}
		\label{cdf2}
\end{figure*}

\begin{figure*}
	\centering
		\includegraphics*[scale=0.74]{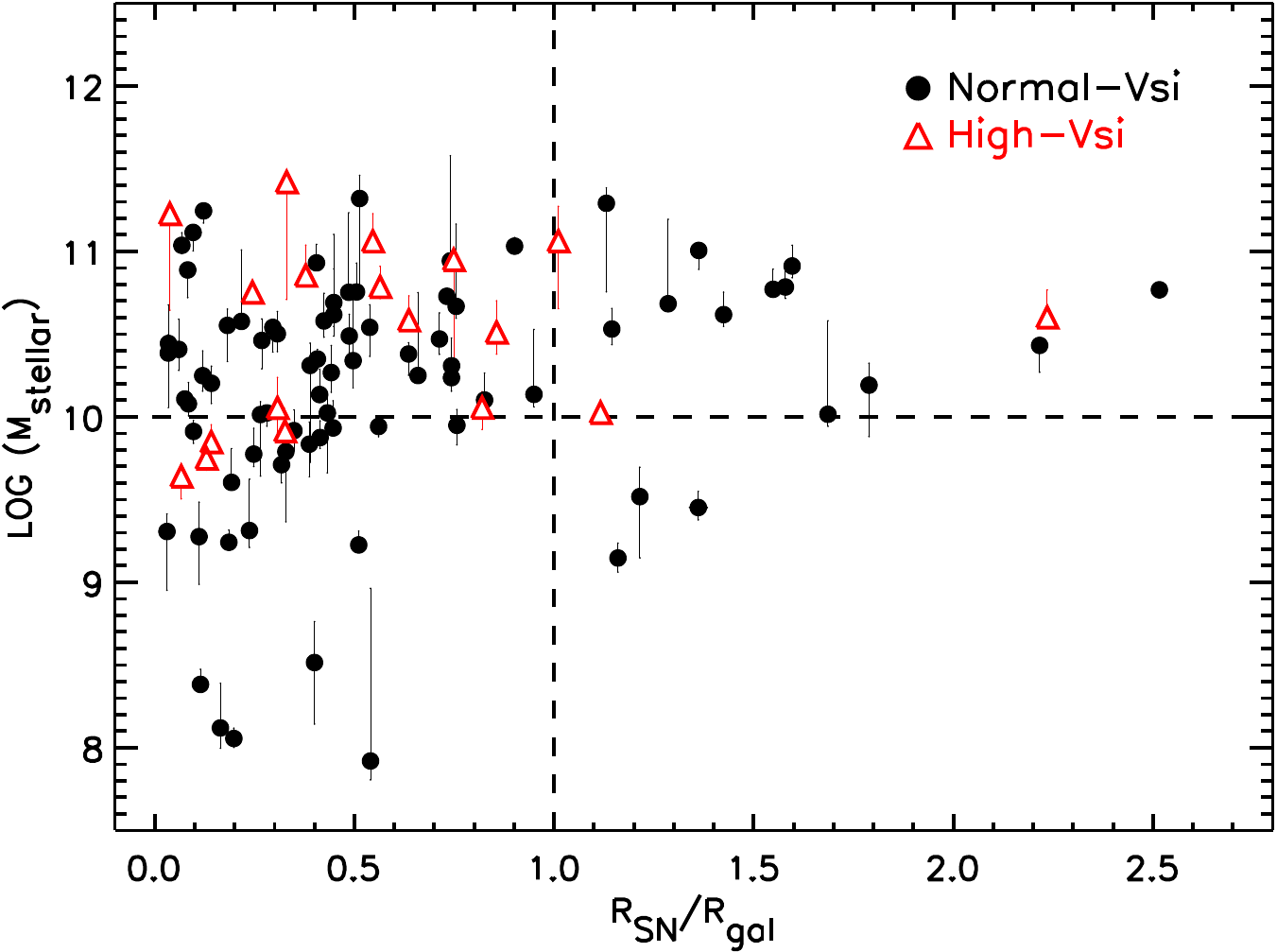}
                \caption{The host galaxy \mstellar\ as a function of
                  \offset. The data are colour-coded in terms of the
                  \Siii\ velocity (\vsiii): high-\vsiii\ SNe Ia (red)
                  and normal-\vsiii\ SNe Ia (black).  }
        \label{offset-mass}
\end{figure*}

\begin{figure}
	\centering
		\includegraphics*[scale=0.5]{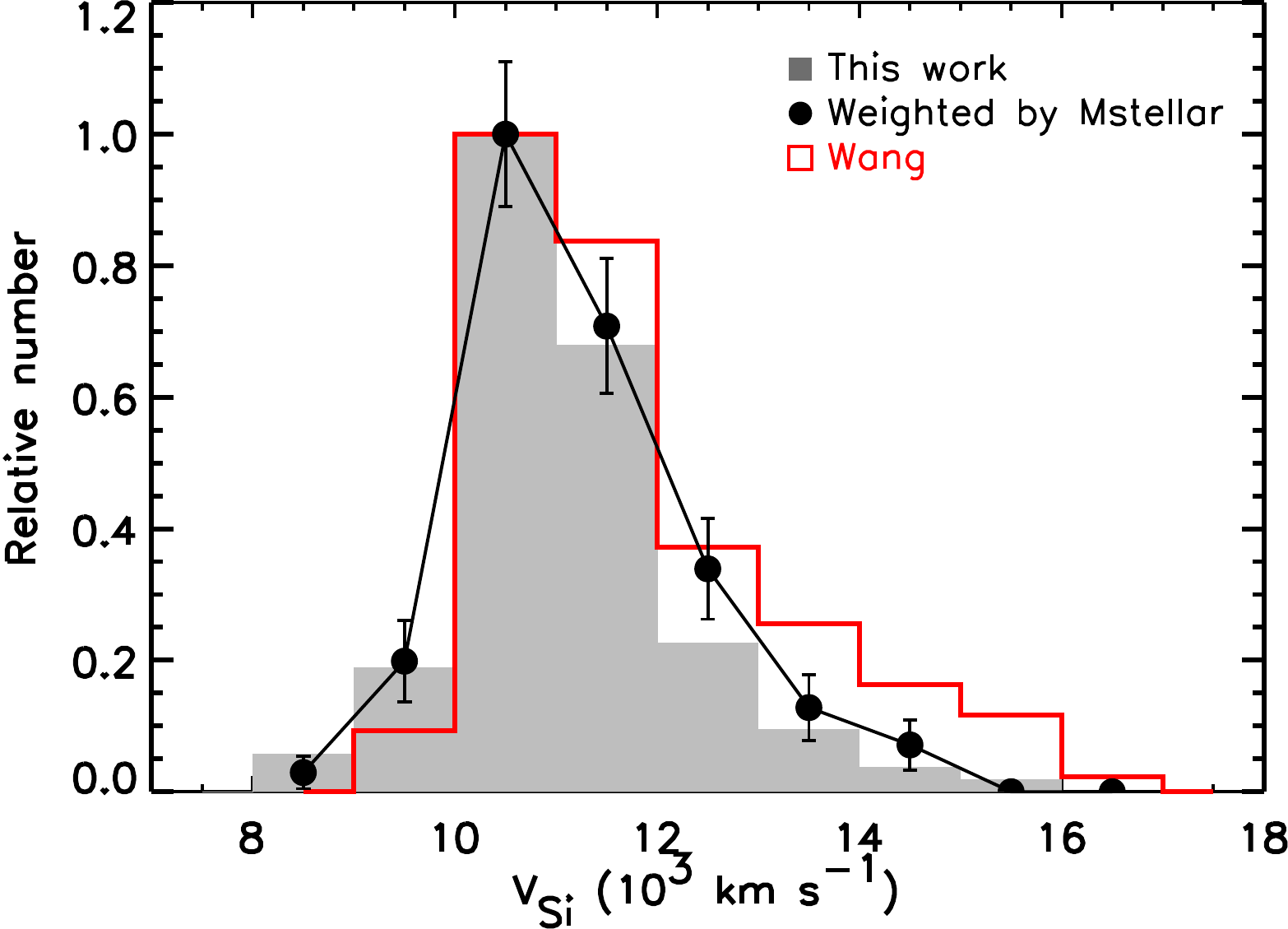}
                \caption{The comparison of the \vsiii\ distributions
                  in the PTF sample (filled histogram) compared with
                  the \citet{2013Sci...340..170W} sample (open red
                  histogram).  The filled circles connected by solid
                  lines show the distribution of our sample weighted
                  by the \mstellar\ of \citet{2013Sci...340..170W},
                  in an attempt to account for the
                  selection differences seen in
                  Fig.~\ref{sample_selection}.  }
        \label{vsi6150_fake}
\end{figure}

%------------------------
\subsubsection{SN offset}
\label{sec:offset}
%------------------------

\citet{2013Sci...340..170W} found evidence for two distinct
populations of SNe Ia with respect to their radial distributions in
the host galaxies.  They found that the high-\vsiii\ SNe Ia are
concentrated in the inner regions of their host galaxies, whereas the
normal-\vsiii\ SNe Ia span a wide range of radial distance.  We next
examine this trend in our sample in Fig.~\ref{offset-vsi}, showing
\vsiii\ as a function of the normalised SN offset, \offset\ (see
Section~\ref{sec:host}). The high-\vsiii\ SNe Ia in the PTF sample
also appear deficient in the outer regions of their hosts compared to
the normal-\vsiii\  SNe Ia, which are found at all radii.  The only
high-\vsiii\ SN at a large radius (PTF09djc;
\vsiii$=13013\,\mathrm{km\,s^{-1}}$, \offset$=2.2$) resides in the
outskirts of an extended galaxy, with no potential host found at the
SN position to the SDSS photometric limit ($r\simeq22\,\rm{mag}$).

However, in our sample the trend is not statistically significant.  We
found that the locations of 83\% of the high-\vsiii\ SNe Ia and 81\%
of the normal-\vsiii\ SNe Ia are within $\offset=1$, which implies the
high-\vsiii\ SNe Ia are similar to the normal-\vsiii\ SNe Ia with
respect to their radial distribution, although the normal-\vsiii\ SN
sample size is much larger.  Fig.~\ref{cdf2} shows the cumulative
distribution functions of \offset\ for high-\vsiii\ and normal-\vsiii\
SNe Ia, and \vsiii\ for SNe at $\offset>1$ and $\offset<1$. We do not
see significant differences in the distributions; a K-S test gives
$p$=0.87 that high-\vsiii\ SNe Ia and normal-\vsiii\ SNe Ia are drawn
from the same population in \offset.  This value is larger than that
for the \citet{2013Sci...340..170W} sample ($p=0.005$).  Repeating the
same analysis as Section~\ref{sec:mass} and randomly selecting 18 SNe
from the full sample, in only 3.3\% of cases do all the selected SNe
have \offset\ smaller than 1.1 (the maximum \offset\ of the
high-\vsiii\ SNe Ia).  However, this test excludes the outlier
PTF09djc.  We see very similar results when including all PTF SNe Ia
up to $z<0.15$, and similar results when only considering SNe
  Ia in massive host galaxies. We thus conclude that there is no
strong evidence in our sample that high-\vsiii\ SNe Ia prefer the
inner regions of their host galaxies.

Fig.~\ref{offset-mass} shows the host \mstellar\ as a function
  of \offset. The data points are colour-coded in terms of the \Siii\
  velocities. This diagram provides us different angle to investigate
  the properties of the high-\vsiii\ SNe Ia. The high-\vsiii\ SNe Ia
  tend to occupy a parameter space which has both higher \mstellar\
  and smaller R${_\mathrm{SN}}$/R${_\mathrm{gal}}$, consistent with
  our earlier results.

One explanation for the differing level of significance, despite the
similar overall sizes of this and the \citet{2013Sci...340..170W}
sample, is that the PTF sample has a smaller fraction of high-\vsiii\
SNe Ia (20\%; 20 high-\vsiii\ SNe and 102 normal-\vsiii\ SNe) compared
to \citet{2013Sci...340..170W} (33\%; 40 high-\vsiii\ SNe Ia and 83
normal-\vsiii\ SNe Ia).  This may be because the
\citet{2013Sci...340..170W} sample is biased towards both massive
galaxies and higher surface brightness galaxies
(Fig.~\ref{sample_selection} and Fig.~\ref{SB_compare}; Section~\ref{sec:compare-other-sample}),
whereas PTF is untargeted.  If high-\vsiii\ SNe Ia occur more
frequently in massive galaxies (e.g., Fig.~\ref{m-vsi};
Section~\ref{sec:mass}), we would expect more high-\vsiii\ SNe Ia to
be discovered in a galaxy-targeted survey such as LOSS.

We test this hypothesis by attempting to reproduce the \vsiii\
distribution of \citet{2013Sci...340..170W} using the PTF sample, by
matching the \citeauthor{2013Sci...340..170W} selection in \mstellar.
We generate synthetic SN samples by selecting 123 events (the size of
the \citeauthor{2013Sci...340..170W} sample) from the PTF sample, with
the probability of a SN being selected weighted by its \mstellar,
using the \mstellar\ distribution from \citet{2013Sci...340..170W}.
We repeated this procedure 10,000 times and determined the mean
\vsiii\ distribution.  The result is shown in Fig.~\ref{vsi6150_fake}.

Weighting by \mstellar\ clearly alters the \vsiii\ distribution of our
synthetic sample, as expected. The distribution in \vsiii\ between
8000 and 13000\,$\mathrm{km\,s^{-1}}$, after weighting by \mstellar,
is now consistent with \citet{2013Sci...340..170W}. However, the
distribution is still inconsistent at
$\vsiii>13000$\,$\mathrm{km\,s^{-1}}$. Thus although the selection
bias in \mstellar\ may explain some of the difference, it is clear
that either some other variable is also at work, or our tests of the
selection effects do not capture all the effect.

%------------------------------------
\subsubsection{Other host parameters}
\label{sec:other-host}
%------------------------------------
\begin{figure*}
	\centering
	\includegraphics*[scale=0.7]{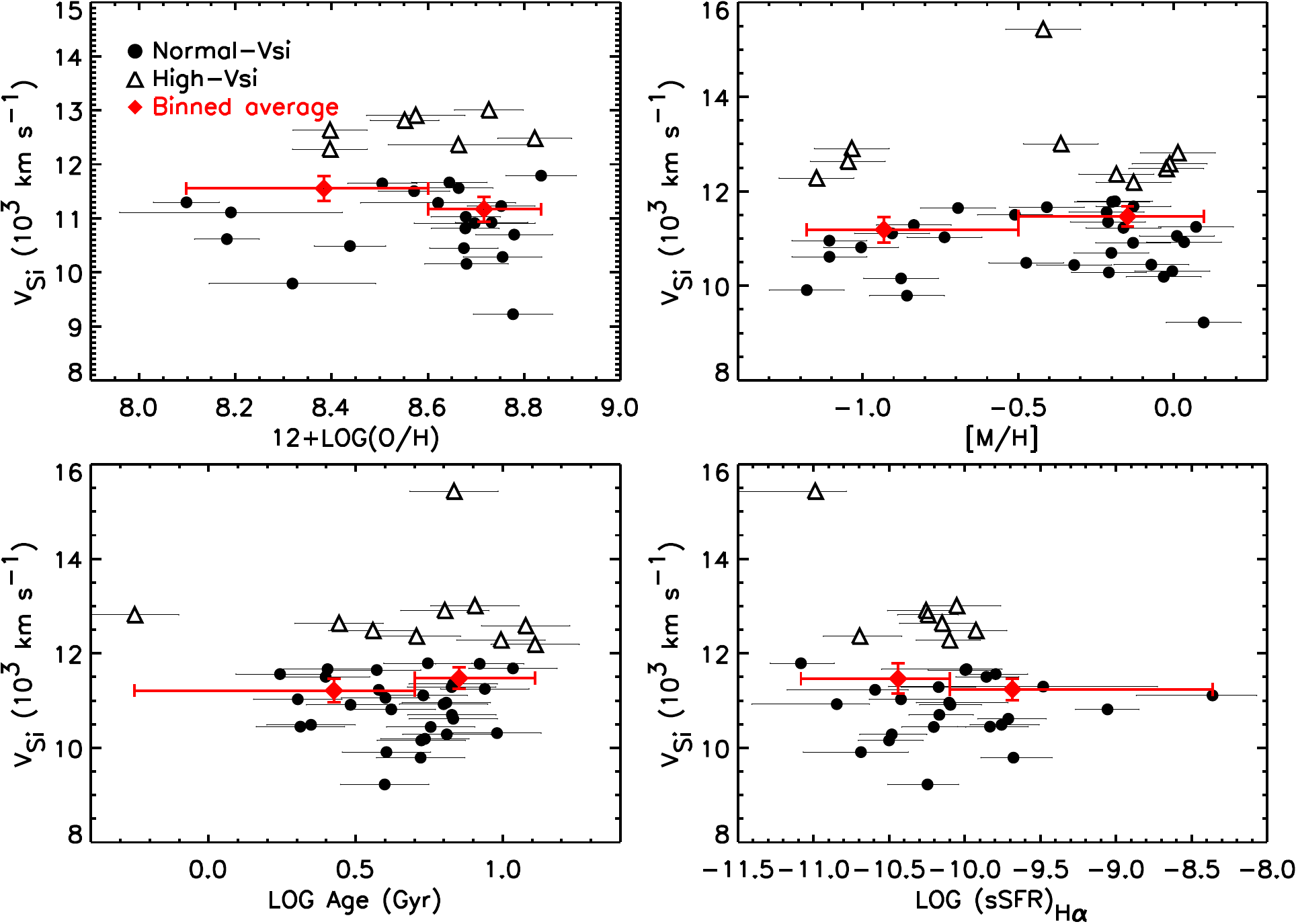}
	\caption{The \Siii\ velocity (\vsiii) as a function of
          gas-phase metallicity (upper left), stellar metallicity
          (upper right), stellar age (lower left) and specific
          star-formation rate (lower right). The red diamonds
          represent the mean of \vsiii\ in bins of each of the host
          parameters, and the error bars are the width of the bins and
          the error of the mean.}
	\label{host-vsi6150}
\end{figure*}

As discussed in Section~\ref{sec:host}, some of the host galaxies in
our sample also have host spectral parameters measured in
\citet{2014MNRAS.438.1391P}. These include specific star formation
rate (sSFR), gas-phase metallicity, stellar metallicity and stellar
age, and in Fig.~\ref{host-vsi6150}, we present the \vsiii\ as a
function of these parameters. The sample size is smaller than for the
investigations related to \mstellar\ or \offset, and we do not find
any significant trends between \vsiii\ and these host properties.

%=================================
\subsection{High-velocity features and host parameters}
\label{sec:hvfs-host}
%=================================
\begin{figure*}
	\centering
	\includegraphics*[scale=0.7]{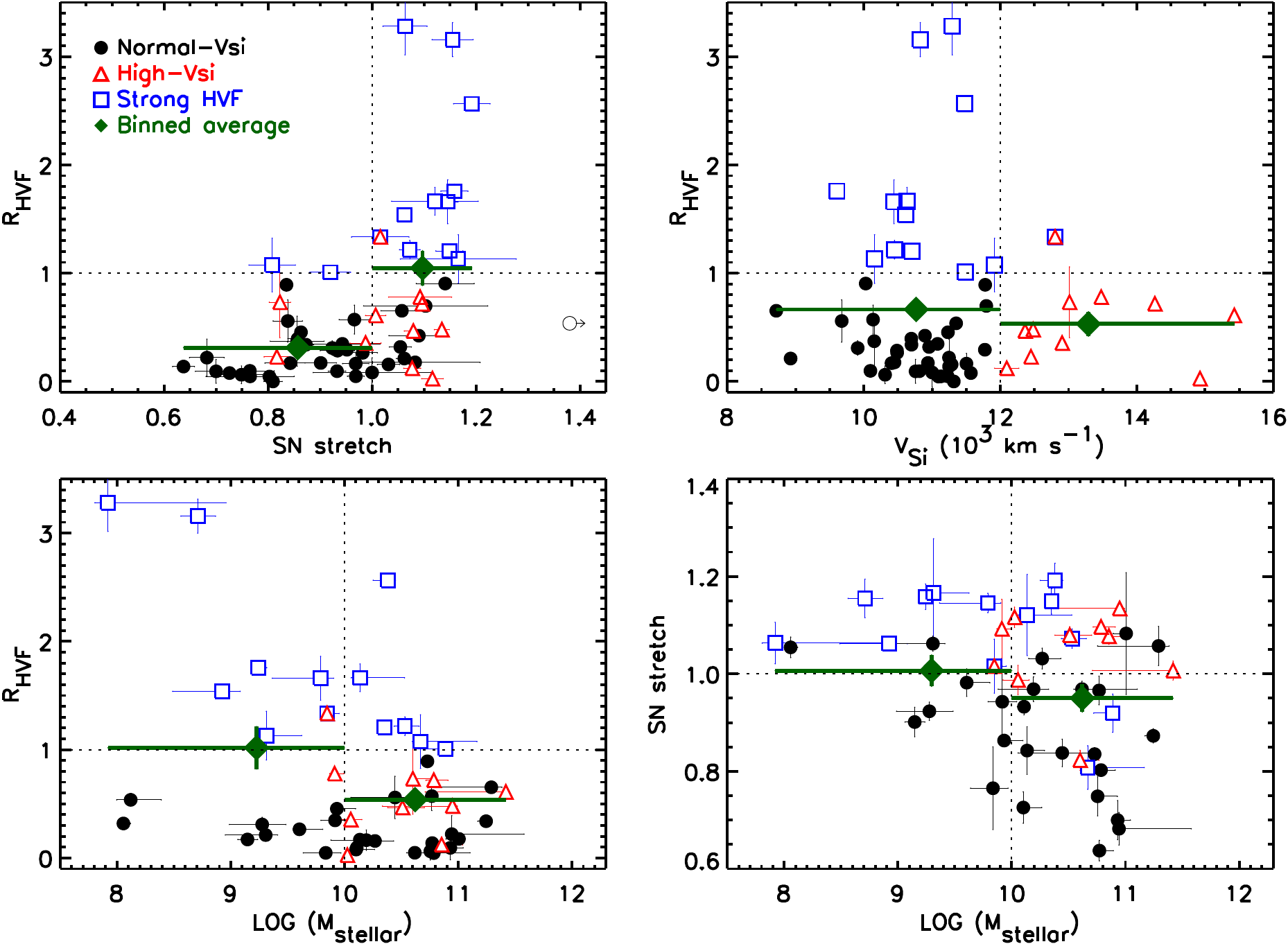}
	\caption{The ratio of the pEW of the high-velocity \Caii\ NIR
          feature to the photospheric \Caii\ feature, \rhvf\ as a
          function of various parameters. Upper left is the SN
          stretch, upper right the \Siii\ velocity, and lower left the
          \mstellar.  The SN stretch is plotted against \mstellar\ in
          the lower-right panel. Only SNe with \Caii\ NIR features are
          plotted.  The solid circles, open triangles and open squares
          represent the normal-\vsiii\ SNe Ia, high-\vsiii\ SNe Ia and
          SNe Ia with a large \rhvf) respectively.  The solid green
          diamonds represent the mean \rhvf\ in each bin.  The open
          circle shows one outlier in the plot: PTF09dhx ($s=1.7$,
          $\rhvf=0.54$) }
	\label{rhvf}
\end{figure*}

\begin{figure*}
	\centering
	\includegraphics*[scale=0.7]{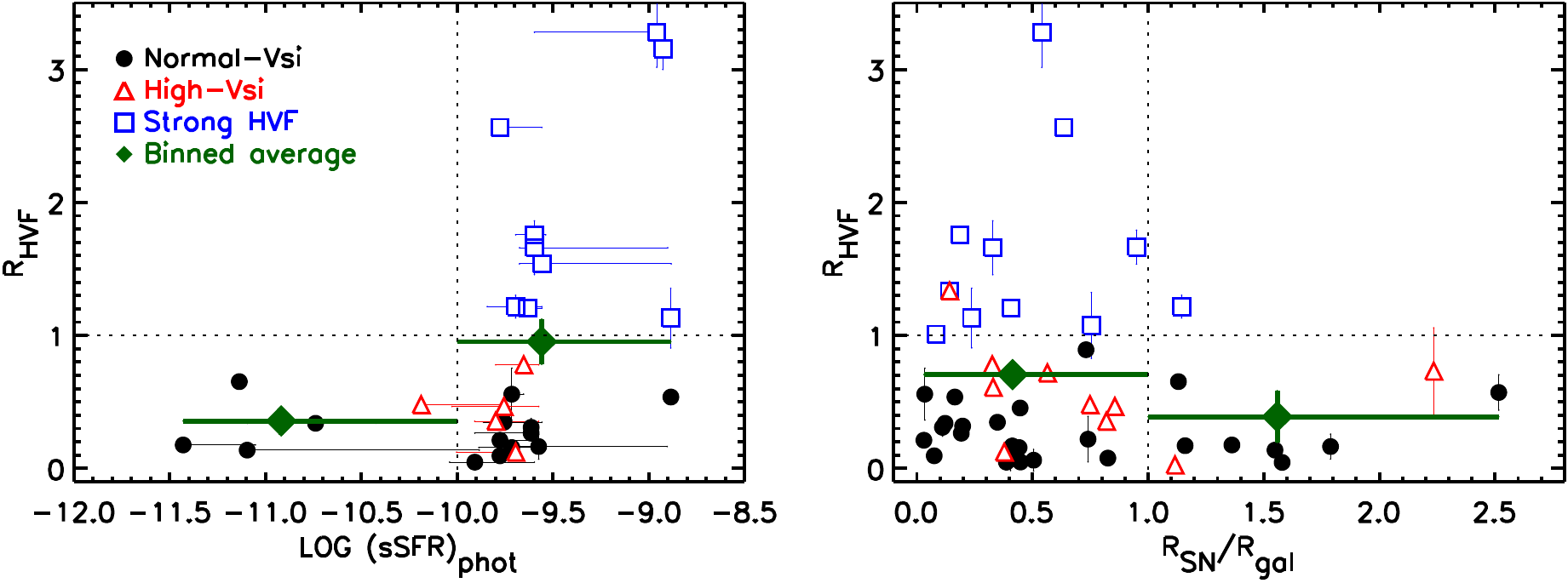}
	\caption{The same as Fig.~\ref{rhvf} but with the sSFR derived from \textsc{z-peg} (left)
	and \offset\ (right).}
	\label{rhvf2}
\end{figure*}

We now discuss the HVFs seen in the \Caii\ NIR feature. Following the
procedure in \citet{2014MNRAS.437..338C}, we quantify the strength of
the HVFs using the ratio (\rhvf) of the pEWs of the \Caii\ NIR
high-velocity component to the pEW of the photospheric component.  SNe
Ia with a larger \rhvf\ will present stronger relative absorption in
the high-velocity component compared to the photospheric component.
We define SNe Ia with $\rhvf>1.0$ as having strong HVFs.

Fig.~\ref{rhvf} shows \rhvf\ as a function of the SN stretch ($s$),
\vsiii, and \mstellar\ \citep[see][for the correlation between \rhvf\
and stretch for the full PTF sample]{2014MNRAS.444.3258M}.  We see a clear
trend that nearly all SNe Ia with a large \rhvf\ have high stretches
($s\ga1.0$), as expected based on earlier work
\citep{2014MNRAS.437..338C}. This trend could be driven either by
stronger HVFs in high-stretch SNe Ia, or by weaker photospheric
features.  \citet{2014MNRAS.444.3258M} demonstrated that both effects are
present: high-stretch SNe have both weaker photospheric features, and
stronger HVFs. The former can be understood if high stretch SNe Ia
have higher temperatures, and therefor a higher ionisation:
photospheric \Caii\ is ionised to \Caiii\ and thus the \Caii\ pEW
becomes weaker.

There is one outlier in Fig.~\ref{rhvf} with very high stretch but
weak HVF: PTF09dhx ($s=1.7$, $\rhvf=0.54$).  By contrast, SNe Ia with
high-\vsiii\ have relatively weak HVFs and intermediate stretches
($0.8<s<1.1$).  We also confirm that SNe Ia with a large \rhvf\ have
normal \Siii\ velocities (\vsiii\ $<12000\,\mathrm{km\,s^{-1}}$),
already noted by \citet{2014MNRAS.437..338C}. There is only one
SN in our sample has both strong HVFs and high \Siii\ velocity 
(PTF10lot; \vsiii$=12802\,\mathrm{km\,s^{-1}}$ and \rhvf$=1.33$).
The lower panels of Fig.~\ref{rhvf} show the relations between SN stretch, \mstellar\ and
\rhvf.  In contrast to the high-\vsiii\ SNe Ia, which are likely to
reside in massive galaxies, here we find that SNe Ia with a large
\rhvf\ are preferentially found in low-mass galaxies. Thus SNe Ia with
high-\vsiii\ appear different in terms of their host galaxies to SNe
Ia with high-velocity \Caii\ NIR features.

In Fig.~\ref{rhvf2} we examine \rhvf\ as a function of the specific
star-formation rate (sSFR) and \offset.  Here we use the sSFR
determined by \textsc{z-peg}, instead of using the H${\alpha}$
luminosity, to increase the sample size. The events with a large
\rhvf\ all reside in strongly star-forming galaxies, and indeed nearly
all SNe Ia in such galaxies display strong HVFs. By contrast, SNe Ia
with high-\vsiii\ tend to reside in galaxies with intermediate sSFRs:
$\log(\mathrm{sSFR})\sim-9.7$.  For the relation between \rhvf\ and
\offset, we found all SNe Ia with a large \rhvf\ are located within
$\offset=1$ of their host galaxies.  As we will see in
Section~\ref{sec:origin-of-hvfs}, these HVFs mostly come from
late-type galaxies (spirals or irregulars).  This strongly implies
that SNe Ia with HVFs are less likely to originate from old
populations residing in galactic haloes distant from the host centre.

%%%%%%%%%%%%%%%%%%%%%%%%%%%%%%%%%%%%%%%%%%%%%%%%%%%%%%%%%%%%%%%%%%%%%%
\section{Discussion}
\label{sec:discussion}
%%%%%%%%%%%%%%%%%%%%%%%%%%%%%%%%%%%%%%%%%%%%%%%%%%%%%%%%%%%%%%%%%%%%%%

%-----------------------------
\subsection{Silicon velocity and metallicity}
\label{sec:vsi}
%-----------------------------
\citet{2013Sci...340..170W} found that high-\vsiii\ SNe Ia appear
concentrated in the inner regions of their host galaxies, whereas
normal-\vsiii\ SNe Ia span a wider range of radial distance.
Observations have shown that negative metallicity gradients are common
in both the Milky Way and many external galaxies, in the sense that
the heavy-element abundances decrease systematically outward from the
centre of galaxies \citep{1999PASP..111..919H}.  Therefore we would
expect these high-\vsiii\ SNe Ia are more likely to originate from
metal-rich (and older) populations. In this work we did not find any
significant trends between high-\vsiii\ SNe Ia and radial position.
However, we did observe a stronger trend that high-\vsiii\ SNe Ia tend
to explode in more massive galaxies. According to the galaxy
mass-metallicity relation
\citep{2004ApJ...613..898T,2008ApJ...681.1183K}, massive galaxies are
generally more metal-rich than low-mass galaxies, and thus the stellar
populations in the inner regions of galaxies may be similar to massive
galaxies with respect to their metallicities. By using the \mstellar\
as a different approach, we suspect metallicity may be a potential
variable in making high-\vsiii\ SNe Ia different from normal-\vsiii\
SNe Ia. This is also supported by the evidence that high-\vsiii\ SNe
Ia appear to be redder than normal-\vsiii\ SNe Ia
\citep{2011ApJ...729...55F}, together with recent observational
results that SNe Ia in metal-rich environments are redder than those
in metal-poor environments
\citep{2013ApJ...770..108C,2014MNRAS.438.1391P}.

\citet{2000ApJ...530..966L} showed that the C+O layer metallicity in
SN Ia explosion could play a role in affecting the observed \Siii\
velocity. The blue-shifted velocities of the silicon features increase
with C+O layer metallicity due to the increasing opacity in the C+O
layer moving the features blueward and causing larger line velocities.
We determined a linear relation between the \Siii\ velocity and C+O
metallicity of SN progenitor using the models of
\citet{2000ApJ...530..966L}. The \Siii\ velocities were measured from
their SN Ia model spectra using the same technique described in
Section~\ref{sec:spectral-measurements}. The result shows the \Siii\
velocities increase with metallicities with a slope of
$\sim$435\,km\,$\rm s^{-1}$\,$\rm dex^{-1}$.

We determined the trend between \Siii\ velocity and metallicity using
our sample by converting our host \mstellar\ to gas-phase metallicity
in Fig.~\ref{m-vsi}.  The mass--metallicity relation studied from
\citet{2008ApJ...681.1183K} was used for the conversion.  We found the
\Siii\ velocities increase with metallicity with a slope of
$\sim$1357\,km\,$\rm s^{-1}$\,$\rm dex^{-1}$.  The slope determined
from our data is in qualitative agreement with that of
\citet{2000ApJ...530..966L} models, although ours shows a steeper
trend.  The number of direct metallicity measurements from host
spectroscopic data in this work do not allow for a sufficient
statistical power to reveal if metallicity is important factor in
altering the \vsiii.  However, given the tight relation between
\mstellar\ and metallicity, this does offer a possible explanation
that high-\vsiii\ SNe Ia could originate from more metal-rich
populations than normal-\vsiii\ SNe Ia.

%---------------------------------------
\subsection{The physical origin of HVFs}
\label{sec:origin-of-hvfs}
%---------------------------------------

\begin{figure}
	\centering
	\includegraphics*[scale=0.5]{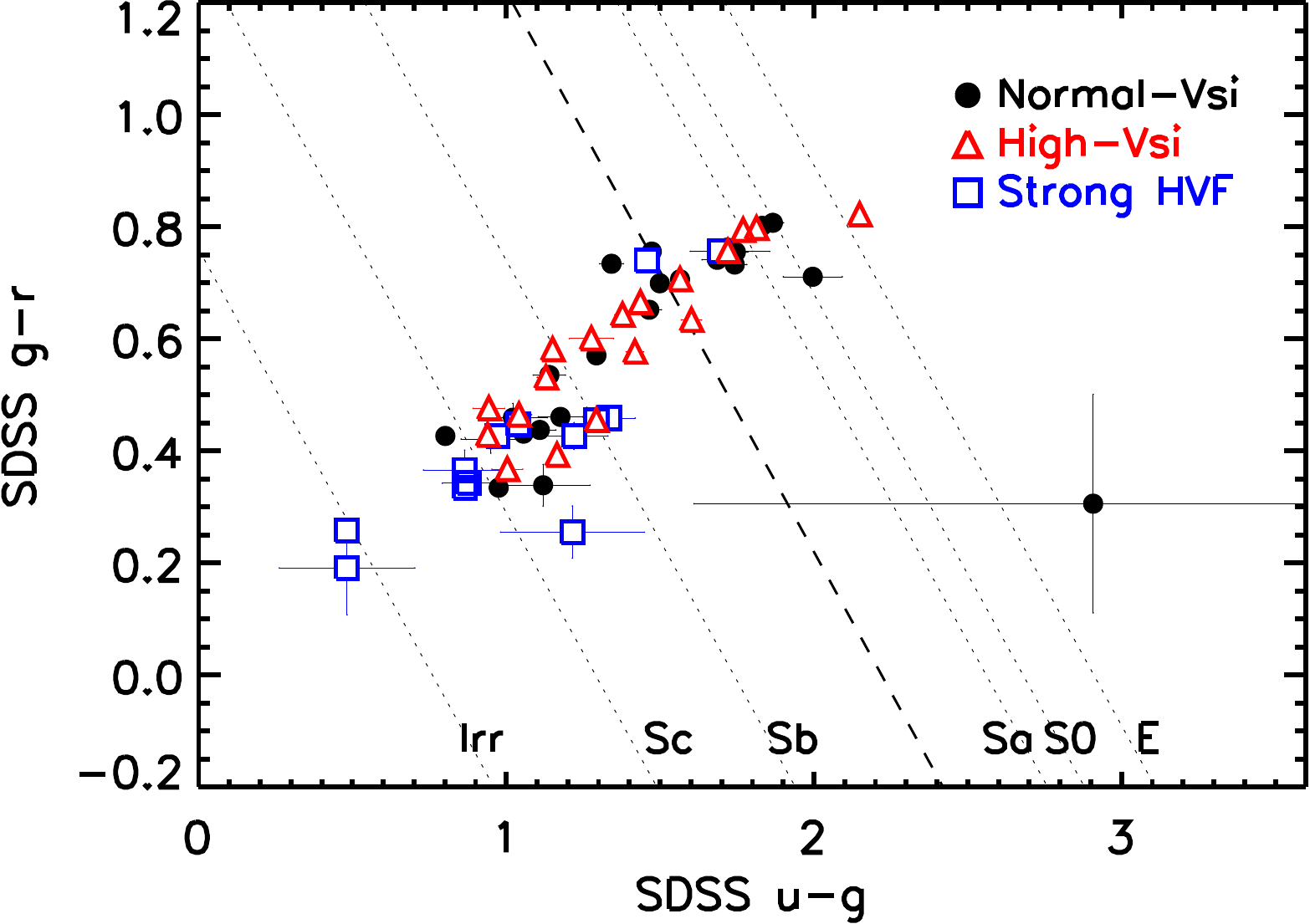}
	\caption{The rest-frame SDSS $g-r$ versus $u-g$ host galaxy colour for
          our SN Ia sample.  The empirical relation studied by
          \citet{2001AJ....122.1861S} is used to type the host
          galaxies.  The dotted lines show the average $u-r$ for
          Irr(0.76), Sc(1.29), Sb(1.74), Sa(2.56), S0(2.68) and
          E(2.91) galaxies and the dashed line is the criterion to
          separate the late-type (left) and early-type (right)
          galaxies. The normal-\vsiii\ SNe Ia, high-\vsiii\ SNe Ia and
          SNe Ia with strong HVFs are shown in solid circles, open
          triangles and open squares, respectively.}
	\label{color-color}
\end{figure}

\begin{table*}
\centering
\caption{The comparison between normal-\vsiii/weak HVFs SNe Ia, high-\vsiii\ SNe Ia and SNe Ia with strong HVFs.}
\begin{tabular}{lccc}
  \hline\hline
& Normal \vsiii\  & High \vsiii\ & Strong HVFs   \\
& \& weak HVFs    &              &               \\                
  \hline
  SN stretch & $0.6<s<1.1$ &$0.8<s<1.1$ & $s>1.0$  \\
  \vsiii\ & -- & -- & $<12000$ km\,$\mathrm{s^{-1}}$  \\
  Host type & E to Sc & E to Sc & Sb/Sc/Irr  \\
  Host \mstellar\ & low mass/massive & massive & prefer low mass \\
  Host sSFR & low/intermediate & intermediate & high \\
  \hline
\end{tabular}
\label{hv-hvfs}
\end{table*}

In this work, we found SNe Ia with strong \Caii\ NIR HVFs present
different properties to those with weaker HVFs.  The physical origin
of HVFs in SNe Ia is not yet clear.  They are common in SN Ia spectra,
and many scenarios have been proposed by previous studies
\citep{2003ApJ...591.1110W,2004ApJ...607..391G,2005ApJ...623L..37M,2005MNRAS.357..200M,2006ApJ...636..400Q, 2006ApJ...645..470T,2008ApJ...677..448T,2013ApJ...770...29C,2014MNRAS.437..338C}.
Briefly speaking, the HVFs could be produced by either an abundance
enhancement or a density enhancement.
An abundance enhancement could be caused by an enhancement of intermediate-mass elements (IMEs) in the outer regions of the SN ejecta. 
A mixing with hydrogen from circumstellar 
material (CSM) could also increase the electron density and strengthen the recombination, 
which will result in a stronger \Caii\ feature.
A density enhancement could result from the SN
explosion itself, or the interaction between SN ejecta and
CSM -- or both .

We have confirmed that SNe Ia with a large \rhvf\ are higher stretch
(and therefore brighter) than those with a small \rhvf.  We found a
strong trend that they are found in galaxies with a lower stellar mass
and a stronger sSFR. \citet{2014MNRAS.444.3258M} show that although this
relation is partially driven by higher-stretch SNe Ia having weaker
photospheric \Caii\ NIR, this does not explain the entire trend;
higher stretch SNe Ia also have stronger HVFs.

In Fig.~\ref{color-color} we plot the rest-frame
colour-colour diagram (SDSS $g-r$ against $u-g$) of our host galaxies.
Following the procedure described in \citet{2006A&A...448..893L}, we
further classify our host galaxies into different Hubble types based
on the criteria proposed by \citet{2001AJ....122.1861S}. The mean
$u-r$ values for six different spectral types of galaxies are
overplotted.  Although the $u-r$ values used here have some
uncertainty due to dust extinction in the galaxy, overall it provides
a good idea of the Hubble types of our hosts.

The host galaxies of SNe Ia with a large \rhvf\ concentrate toward the
blue end of the host galaxy colour sequence. These galaxies are mostly
classified as Sb/Sc/Irr galaxies, whereas the high-\vsiii\ SNe Ia are
found in both late-type and early-type galaxies. This is consistent
with Section~\ref{sec:hvfs-host}, where we showed that SNe Ia with
HVFs arise in galaxies with very strong sSFRs, and are therefore likely
to be related to young stellar populations. This argues against an
orientation or viewing angle effect
\citep[confirming the results of][]{2006ApJ...645..470T} being purely responsible for the
presence of HVFs in SN Ia spectra, as there is no reason that the
orientation would depend on the underlying stellar population.

Using narrow blue-shifted Na\,\textsc{i} D absorption features as a
probe of this CSM, \citet{2011Sci...333..856S} and
\citet{2013MNRAS.436..222M} found an excess of SNe Ia with blueshifted
narrow Na\,\textsc{i} D features, showing CSM around their
progenitors.  They further found the host galaxies of these SNe Ia are
mostly late-type galaxies. If the HVFs observed in SN spectra are
related to the interaction between the SN ejecta and a CSM, our work
is consistent with \citeauthor{2013MNRAS.436..222M}: SNe Ia presenting
strong HVFs tend to explode in galaxies with young stellar
populations. This provides further evidence for at least two different
populations of SNe Ia \citep[see discussion
in][]{2013MNRAS.436..222M}, given the distinct properties of host
galaxies between SNe Ia with strong HVFs and weak HVFs/high-\vsiii\
SNe Ia. In Table~\ref{hv-hvfs} we summarise the properties of
high-\vsiii\ SNe Ia and SNe Ia with strong HVFs found in this work.

%%%%%%%%%%%%%%%%%%%%%%%%%%%%%%%%%%%%%%%%%%%%%%%%%%%%%%%%%%%%%%%%%%%%%%
\section{Conclusions}
\label{sec:conclusions}
%%%%%%%%%%%%%%%%%%%%%%%%%%%%%%%%%%%%%%%%%%%%%%%%%%%%%%%%%%%%%%%%%%%%%%

In this paper we have analysed spectroscopic measurements of 122 type
Ia supernovae (SNe Ia) with $z<0.09$ discovered by the Palomar
Transient Factory. In particular, we focused on the velocity and
pseudo equivalent widths of the \Siii\ and \Caii\ near-infrared
triplet (NIR) absorptions.  We determined the host parameters using
both photometric and spectroscopic data, and estimated the host
\mstellar, SN--host offset, star formation rate (SFR), metallicity and
age.  Various relations between SN spectral features and host
parameters are demonstrated in this work.  Below we summarise our main
findings.

\begin{enumerate}
\item[$\bullet$] We find that SNe Ia with a high-\Siii\ velocity
  (high-\vsiii) are preferentially found in massive host galaxies,
  whereas SNe Ia with normal-\vsiii\ are found in hosts of all stellar
  mass (Fig.~\ref{m-vsi}). We find weaker, but consistent, trends when
  considering the \Caii\ NIR feature in place of \Siii.

\item[$\bullet$] There is also some evidence that these high-\vsiii\
  SNe Ia are found in the inner regions of their host galaxies
  (Fig.~\ref{offset-vsi}).  However, this trend is not as statistically
  significant in our sample as in previous studies.

\item[$\bullet$] If stellar mass is intepreted as a proxy for
  metallicity, and inner regions of galaxies are more metal rich,
  these findings are consistent with a metallicity dependence in
  \vsiii. Such a qualitative dependence is seen in some models of SN
  Ia spectra.

\item[$\bullet$] SNe Ia with a strong \rhvf\ (defined as the ratio of
  the pEWs of the high-velocity and photospheric components of the
  \Caii\ near-IR feature) are preferentially found in galaxies with a
  lower stellar mass, a bluer colour and a stronger sSFR
  (Figs.~\ref{rhvf} and \ref{rhvf2}). Their host SEDs are consistent
  with being of morphological type Sc and later
  (Fig.~\ref{color-color}).

\item[$\bullet$] This strongly suggests that SNe Ia with a large
  \rhvf\ originate from young stellar populations, and argues against
  an orientation effect being purely responsible for HVFs in SNe Ia.
  Previous studies proposed a strong link between SN Ia HVFs and
  circumstellar material (CSM), and found most of the SNe Ia showing
  signatures of CSM explode in late-type galaxies.  Our results are
  consistent with these findings, assuming that the HVFs are related
  to the interaction between the SN ejecta and a CSM local to the SN.

\end{enumerate}

Investigating the relationships between SN spectral features and host
galaxy parameters provides a different angle to probe the nature of
the SN explosion. Host studies have been proven to be useful in
previous works, especially in the relations with SN luminosities.
Here in this work we show there is also a strong connection between SN
Ia spectral features and their host galaxies, which is worth further
investigation and study.

\section*{Acknowledgements}

MS acknowledges support from the Royal Society and EU/FP7-ERC grant no
[615929].  A.G.-Y. is supported by the EU/FP7-ERC grant no [307260],
the Quantum Universe I-Core program by the Israeli Committee for
planning and funding and the ISF, GIF, Minerva and ISF grants, and
Kimmel and ARCHES awards.  This research used resources of the National 
Energy Research Scientific Computing Center, which is supported by the 
Office of Science of the U.S. Department of Energy under Contract 
No. DE-AC02-05CH11231. Observations obtained with the Samuel
Oschin Telescope at the Palomar Observatory as part of the Palomar
Transient Factory project, a scientific collaboration between the
California Institute of Technology, Columbia University, Las Cumbres
Observatory, the Lawrence Berkeley National Laboratory, the National
Energy Research Scientific Computing Center, the University of Oxford,
and the Weizmann Institute of Science.  The William Herschel Telescope
is operated on the island of La Palma by the Isaac Newton Group in the
Spanish Observatorio del Roque de los Muchachos of the Instituto de
Astrofisica de Canarias.  Based on observations obtained at the Gemini
Observatory, which is operated by the Association of Universities for
Research in Astronomy, Inc., under a cooperative agreement with the
NSF on behalf of the Gemini partnership: the National Science
Foundation (United States), the National Research Council (Canada),
CONICYT (Chile), the Australian Research Council (Australia),
Minist\'{e}rio da Ci\^{e}ncia, Tecnologia e Inova\c{c}\~{a}o (Brazil)
and Ministerio de Ciencia, Tecnolog\'{i}a e Innovaci\'{o}n Productiva
(Argentina). Based on Gemini programs GN-2010B-Q-111, GS-2010B-Q-82,
GN-2011A-Q-82, GN-2011B-Q-108, GN-2012A-Q-91, GS-2012A-Q-3,
GN-2012B-Q-122, and GS-2012B-Q-83 for the host galaxy observations,
and GN-2010A-Q-20, GN-2010B-Q-13, GN-2011A-Q-16 and GS-2009B-Q-11 for
the SN observations.  This work makes use of observations from the
LCOGT network.  Some of the data presented herein were obtained at the
W.M. Keck Observatory, which is operated as a scientific partnership
among the California Institute of Technology, the University of
California and the National Aeronautics and Space Administration. The
Observatory was made possible by the generous financial support of the
W.M. Keck Foundation.  Based on observations collected at the European
Organisation for Astronomical Research in the Southern Hemisphere,
Chile, under program IDs 084.A-0149 and 085.A-0777.  Observations
obtained with the SuperNova Integral Field Spectrograph on the
University of Hawaii 2.2-m telescope as part of the Nearby Supernova
Factory II project, a scientific collaboration between the Centre de
Recherche Astronomique de Lyon, Institut de Physique Nucl'eaire de
Lyon, Laboratoire de Physique Nucl'eaire et des Hautes Energies,
Lawrence Berkeley National Laboratory, Yale University, University of
Bonn, Max Planck Institute for Astrophysics, Tsinghua Center for
Astrophysics, and Centre de Physique des Particules de Marseille.

This research has made use of the NASA/IPAC Extragalactic Database
(NED) which is operated by the Jet Propulsion Laboratory, California
Institute of Technology, under contract with the National Aeronautics
and Space Administration.

This publication has been made possible by the participation of more
than 10 000 volunteers in the Galaxy Zoo: Supernovae project
(\texttt{http://supernova.galaxyzoo.org/authors}).

\bibliographystyle{mn2e}
\bibliography{pan2013b}

\begin{thebibliography}{100}
\expandafter\ifx\csname natexlab\endcsname\relax\def\natexlab#1{#1}\fi

\bibitem[{{Ahn} {et~al}\mbox{.}(2013){Ahn}, {Alexandroff}, {Allende Prieto},
  {Anders}, {Anderson}, {Anderton}, {Andrews}, {Aubourg}, {Bailey}, {Bastien},
  \& et~al.}]{2013arXiv1307.7735A}
{Ahn} C.~P. {et~al.}, 2013, ArXiv e-prints

\bibitem[{{Bailey} {et~al}\mbox{.}(2009){Bailey}, {Aldering}, {Antilogus},
  {Aragon}, {Baltay}, {Bongard}, {Buton}, {Childress}, {Chotard}, {Copin},
  {Gangler}, {Loken}, {Nugent}, {Pain}, {Pecontal}, {Pereira}, {Perlmutter},
  {Rabinowitz}, {Rigaudier}, {Ripoche}, {Runge}, {Scalzo}, {Smadja}, {Tao},
  {Thomas}, \& {Wu}}]{2009AAS...21332105B}
{Bailey} S.~J. {et~al.}, 2009, in Bulletin of the American Astronomical
  Society, Vol.~41, American Astronomical Society Meeting Abstracts \#213, p.
  321.05

\bibitem[{{Baldwin} {et~al}\mbox{.}(1981){Baldwin}, {Phillips}, \&
  {Terlevich}}]{1981PASP...93....5B}
{Baldwin} J.~A., {Phillips} M.~M., {Terlevich} R., 1981, \pasp, 93, 5

\bibitem[{{Benetti} {et~al}\mbox{.}(2005){Benetti}, {Cappellaro}, {Mazzali},
  {Turatto}, {Altavilla}, {Bufano}, {Elias-Rosa}, {Kotak}, {Pignata}, {Salvo},
  \& {Stanishev}}]{2005ApJ...623.1011B}
{Benetti} S. {et~al.}, 2005, \apj, 623, 1011

\bibitem[{{Bertin} \& {Arnouts}(1996)}]{1996A&AS..117..393B}
{Bertin} E., {Arnouts} S., 1996, \aaps, 117, 393

\bibitem[{{Betoule} {et~al}\mbox{.}(2014){Betoule}, {Kessler}, {Guy}, {Mosher},
  {Hardin}, {Biswas}, {Astier}, {El-Hage}, {Konig}, {Kuhlmann}, {Marriner},
  {Pain}, {Regnault}, {Balland}, {Bassett}, {Brown}, {Campbell}, {Carlberg},
  {Cellier-Holzem}, {Cinabro}, {Conley}, {D'Andrea}, {DePoy}, {Doi}, {Ellis},
  {Fabbro}, {Filippenko}, {Foley}, {Frieman}, {Fouchez}, {Galbany}, {Goobar},
  {Gupta}, {Hill}, {Hlozek}, {Hogan}, {Hook}, {Howell}, {Jha}, {Le Guillou},
  {Leloudas}, {Lidman}, {Marshall}, {M{\"o}ller}, {Mour{\~a}o}, {Neveu},
  {Nichol}, {Olmstead}, {Palanque-Delabrouille}, {Perlmutter}, {Prieto},
  {Pritchet}, {Richmond}, {Riess}, {Ruhlmann-Kleider}, {Sako}, {Schahmaneche},
  {Schneider}, {Smith}, {Sollerman}, {Sullivan}, {Walton}, \&
  {Wheeler}}]{2014arXiv1401.4064B}
{Betoule} M. {et~al.}, 2014, ArXiv e-prints

\bibitem[{{Blondin} {et~al}\mbox{.}(2011){Blondin}, {Mandel}, \&
  {Kirshner}}]{2011A&A...526A..81B}
{Blondin} S., {Mandel} K.~S., {Kirshner} R.~P., 2011, \aap, 526, A81

\bibitem[{{Blondin} {et~al}\mbox{.}(2012){Blondin}, {Matheson}, {Kirshner},
  {Mandel}, {Berlind}, {Calkins}, {Challis}, {Garnavich}, {Jha}, {Modjaz},
  {Riess}, \& {Schmidt}}]{2012AJ....143..126B}
{Blondin} S. {et~al.}, 2012, \aj, 143, 126

\bibitem[{{Bloom} {et~al}\mbox{.}(2012{\natexlab{a}}){Bloom}, {Kasen}, {Shen},
  {Nugent}, {Butler}, {Graham}, {Howell}, {Kolb}, {Holmes}, {Haswell},
  {Burwitz}, {Rodriguez}, \& {Sullivan}}]{2012ApJ...744L..17B}
{Bloom} J.~S. {et~al.}, 2012{\natexlab{a}}, \apjl, 744, L17

\bibitem[{{Bloom} {et~al}\mbox{.}(2012{\natexlab{b}}){Bloom}, {Richards},
  {Nugent}, {Quimby}, {Kasliwal}, {Starr}, {Poznanski}, {Ofek}, {Cenko},
  {Butler}, {Kulkarni}, {Gal-Yam}, \& {Law}}]{2012PASP..124.1175B}
{Bloom} J.~S. {et~al.}, 2012{\natexlab{b}}, \pasp, 124, 1175

\bibitem[{{Branch} {et~al}\mbox{.}(2006){Branch}, {Dang}, {Hall}, {Ketchum},
  {Melakayil}, {Parrent}, {Troxel}, {Casebeer}, {Jeffery}, \&
  {Baron}}]{2006PASP..118..560B}
{Branch} D. {et~al.}, 2006, \pasp, 118, 560

\bibitem[{{Branch} \& {van den Bergh}(1993)}]{1993AJ....105.2231B}
{Branch} D., {van den Bergh} S., 1993, \aj, 105, 2231

\bibitem[{{Brown} {et~al}\mbox{.}(2013){Brown}, {Baliber}, {Bianco}, {Bowman},
  {Burleson}, {Conway}, {Crellin}, {Depagne}, {De Vera}, {Dilday}, {Dragomir},
  {Dubberley}, {Eastman}, {Elphick}, {Falarski}, {Foale}, {Ford}, {Fulton},
  {Garza}, {Gomez}, {Graham}, {Greene}, {Haldeman}, {Hawkins}, {Haworth},
  {Haynes}, {Hidas}, {Hjelstrom}, {Howell}, {Hygelund}, {Lister}, {Lobdill},
  {Martinez}, {Mullins}, {Norbury}, {Parrent}, {Paulson}, {Petry}, {Pickles},
  {Posner}, {Rosing}, {Ross}, {Sand}, {Saunders}, {Shobbrook}, {Shporer},
  {Street}, {Thomas}, {Tsapras}, {Tufts}, {Valenti}, {Vander Horst}, {Walker},
  {White}, \& {Willis}}]{2013arXiv1305.2437B}
{Brown} T.~M. {et~al.}, 2013, ArXiv e-prints

\bibitem[{{Cappellari} \& {Emsellem}(2004)}]{2004PASP..116..138C}
{Cappellari} M., {Emsellem} E., 2004, \pasp, 116, 138

\bibitem[{{Childress} {et~al}\mbox{.}(2013{\natexlab{a}}){Childress},
  {Aldering}, {Antilogus}, {Aragon}, {Bailey}, {Baltay}, {Bongard}, {Buton},
  {Canto}, {Cellier-Holzem}, {Chotard}, {Copin}, {Fakhouri}, {Gangler}, {Guy},
  {Hsiao}, {Kerschhaggl}, {Kim}, {Kowalski}, {Loken}, {Nugent}, {Paech},
  {Pain}, {Pecontal}, {Pereira}, {Perlmutter}, {Rabinowitz}, {Rigault},
  {Runge}, {Scalzo}, {Smadja}, {Tao}, {Thomas}, {Weaver}, \&
  {Wu}}]{2013ApJ...770..108C}
{Childress} M. {et~al.}, 2013{\natexlab{a}}, \apj, 770, 108

\bibitem[{{Childress} {et~al}\mbox{.}(2014){Childress}, {Filippenko},
  {Ganeshalingam}, \& {Schmidt}}]{2014MNRAS.437..338C}
{Childress} M.~J., {Filippenko} A.~V., {Ganeshalingam} M., {Schmidt} B.~P.,
  2014, \mnras, 437, 338

\bibitem[{{Childress} {et~al}\mbox{.}(2013{\natexlab{b}}){Childress}, {Scalzo},
  {Sim}, {Tucker}, {Yuan}, {Schmidt}, {Cenko}, {Silverman}, {Contreras},
  {Hsiao}, {Phillips}, {Morrell}, {Jha}, {McCully}, {Filippenko}, {Anderson},
  {Benetti}, {Bufano}, {de Jaeger}, {Forster}, {Gal-Yam}, {Le Guillou},
  {Maguire}, {Maund}, {Mazzali}, {Pignata}, {Smartt}, {Spyromilio}, {Sullivan},
  {Taddia}, {Valenti}, {Bayliss}, {Bessell}, {Blanc}, {Carson}, {Clubb}, {de
  Burgh-Day}, {Desjardins}, {Fang}, {Fox}, {Gates}, {Ho}, {Keller}, {Kelly},
  {Lidman}, {Loaring}, {Mould}, {Owers}, {Ozbilgen}, {Pei}, {Pickering},
  {Pracy}, {Rich}, {Schaefer}, {Scott}, {Stritzinger}, {Vogt}, \&
  {Zhou}}]{2013ApJ...770...29C}
{Childress} M.~J. {et~al.}, 2013{\natexlab{b}}, \apj, 770, 29

\bibitem[{{Conley} {et~al}\mbox{.}(2008){Conley}, {Sullivan}, {Hsiao}, {Guy},
  {Astier}, {Balam}, {Balland}, {Basa}, {Carlberg}, {Fouchez}, {Hardin},
  {Howell}, {Hook}, {Pain}, {Perrett}, {Pritchet}, \&
  {Regnault}}]{2008ApJ...681..482C}
{Conley} A. {et~al.}, 2008, \apj, 681, 482

\bibitem[{{D'Andrea} {et~al}\mbox{.}(2011){D'Andrea}, {Gupta}, {Sako},
  {Morris}, {Nichol}, {Brown}, {Campbell}, {Olmstead}, {Frieman}, {Garnavich},
  {Jha}, {Kessler}, {Lampeitl}, {Marriner}, {Schneider}, \&
  {Smith}}]{2011ApJ...743..172D}
{D'Andrea} C.~B. {et~al.}, 2011, \apj, 743, 172

\bibitem[{{Dilday} {et~al}\mbox{.}(2012){Dilday}, {Howell}, {Cenko},
  {Silverman}, {Nugent}, {Sullivan}, {Ben-Ami}, {Bildsten}, {Bolte}, {Endl},
  {Filippenko}, {Gnat}, {Horesh}, {Hsiao}, {Kasliwal}, {Kirkman}, {Maguire},
  {Marcy}, {Moore}, {Pan}, {Parrent}, {Podsiadlowski}, {Quimby}, {Sternberg},
  {Suzuki}, {Tytler}, {Xu}, {Bloom}, {Gal-Yam}, {Hook}, {Kulkarni}, {Law},
  {Ofek}, {Polishook}, \& {Poznanski}}]{2012Sci...337..942D}
{Dilday} B. {et~al.}, 2012, Science, 337, 942

\bibitem[{{Faber} {et~al}\mbox{.}(2003){Faber}, {Phillips}, {Kibrick},
  {Alcott}, {Allen}, {Burrous}, {Cantrall}, {Clarke}, {Coil}, {Cowley},
  {Davis}, {Deich}, {Dietsch}, {Gilmore}, {Harper}, {Hilyard}, {Lewis},
  {McVeigh}, {Newman}, {Osborne}, {Schiavon}, {Stover}, {Tucker}, {Wallace},
  {Wei}, {Wirth}, \& {Wright}}]{2003SPIE.4841.1657F}
{Faber} S.~M. {et~al.}, 2003, in Society of Photo-Optical Instrumentation
  Engineers (SPIE) Conference Series, Vol. 4841, Instrument Design and
  Performance for Optical/Infrared Ground-based Telescopes, {Iye} M.,
  {Moorwood} A.~F.~M., eds., pp. 1657--1669

\bibitem[{{Foley}(2012)}]{2012ApJ...748..127F}
{Foley} R.~J., 2012, \apj, 748, 127

\bibitem[{{Foley}(2013)}]{2013MNRAS.435..273F}
{Foley} R.~J., 2013, \mnras, 435, 273

\bibitem[{{Foley} \& {Kasen}(2011)}]{2011ApJ...729...55F}
{Foley} R.~J., {Kasen} D., 2011, \apj, 729, 55

\bibitem[{{Foley} {et~al}\mbox{.}(2012){Foley}, {Simon}, {Burns}, {Gal-Yam},
  {Hamuy}, {Kirshner}, {Morrell}, {Phillips}, {Shields}, \&
  {Sternberg}}]{2012ApJ...752..101F}
{Foley} R.~J. {et~al.}, 2012, \apj, 752, 101

\bibitem[{{Galbany} {et~al}\mbox{.}(2012){Galbany}, {Miquel}, {{\"O}stman},
  {Brown}, {Cinabro}, {D'Andrea}, {Frieman}, {Jha}, {Marriner}, {Nichol},
  {Nordin}, {Olmstead}, {Sako}, {Schneider}, {Smith}, {Sollerman}, {Pan},
  {Snedden}, {Bizyaev}, {Brewington}, {Malanushenko}, {Malanushenko},
  {Oravetz}, {Simmons}, \& {Shelden}}]{2012ApJ...755..125G}
{Galbany} L. {et~al.}, 2012, \apj, 755, 125

\bibitem[{{Gallagher} {et~al}\mbox{.}(2005){Gallagher}, {Garnavich}, {Berlind},
  {Challis}, {Jha}, \& {Kirshner}}]{2005ApJ...634..210G}
{Gallagher} J.~S., {Garnavich} P.~M., {Berlind} P., {Challis} P., {Jha} S.,
  {Kirshner} R.~P., 2005, \apj, 634, 210

\bibitem[{{Gallagher} {et~al}\mbox{.}(2008){Gallagher}, {Garnavich},
  {Caldwell}, {Kirshner}, {Jha}, {Li}, {Ganeshalingam}, \&
  {Filippenko}}]{2008ApJ...685..752G}
{Gallagher} J.~S., {Garnavich} P.~M., {Caldwell} N., {Kirshner} R.~P., {Jha}
  S.~W., {Li} W., {Ganeshalingam} M., {Filippenko} A.~V., 2008, \apj, 685, 752

\bibitem[{{Gerardy} {et~al}\mbox{.}(2004){Gerardy}, {H{\"o}flich}, {Fesen},
  {Marion}, {Nomoto}, {Quimby}, {Schaefer}, {Wang}, \&
  {Wheeler}}]{2004ApJ...607..391G}
{Gerardy} C.~L. {et~al.}, 2004, \apj, 607, 391

\bibitem[{{Hachinger} {et~al}\mbox{.}(2006){Hachinger}, {Mazzali}, \&
  {Benetti}}]{2006MNRAS.370..299H}
{Hachinger} S., {Mazzali} P.~A., {Benetti} S., 2006, \mnras, 370, 299

\bibitem[{{Hakobyan} {et~al}\mbox{.}(2009){Hakobyan}, {Mamon}, {Petrosian},
  {Kunth}, \& {Turatto}}]{2009A&A...508.1259H}
{Hakobyan} A.~A., {Mamon} G.~A., {Petrosian} A.~R., {Kunth} D., {Turatto} M.,
  2009, \aap, 508, 1259

\bibitem[{{Hamuy} {et~al}\mbox{.}(1996){Hamuy}, {Phillips}, {Suntzeff},
  {Schommer}, {Maza}, \& {Aviles}}]{1996AJ....112.2391H}
{Hamuy} M., {Phillips} M.~M., {Suntzeff} N.~B., {Schommer} R.~A., {Maza} J.,
  {Aviles} R., 1996, \aj, 112, 2391

\bibitem[{{Hamuy} {et~al}\mbox{.}(2000){Hamuy}, {Trager}, {Pinto}, {Phillips},
  {Schommer}, {Ivanov}, \& {Suntzeff}}]{2000AJ....120.1479H}
{Hamuy} M., {Trager} S.~C., {Pinto} P.~A., {Phillips} M.~M., {Schommer} R.~A.,
  {Ivanov} V., {Suntzeff} N.~B., 2000, \aj, 120, 1479

\bibitem[{{Hayden} {et~al}\mbox{.}(2013){Hayden}, {Gupta}, {Garnavich},
  {Mannucci}, {Nichol}, \& {Sako}}]{2013ApJ...764..191H}
{Hayden} B.~T., {Gupta} R.~R., {Garnavich} P.~M., {Mannucci} F., {Nichol}
  R.~C., {Sako} M., 2013, \apj, 764, 191

\bibitem[{{Henry} \& {Worthey}(1999)}]{1999PASP..111..919H}
{Henry} R.~B.~C., {Worthey} G., 1999, \pasp, 111, 919

\bibitem[{{Hook} {et~al}\mbox{.}(2004){Hook}, {J{\o}rgensen},
  {Allington-Smith}, {Davies}, {Metcalfe}, {Murowinski}, \&
  {Crampton}}]{2004PASP..116..425H}
{Hook} I.~M., {J{\o}rgensen} I., {Allington-Smith} J.~R., {Davies} R.~L.,
  {Metcalfe} N., {Murowinski} R.~G., {Crampton} D., 2004, \pasp, 116, 425

\bibitem[{{Iben} \& {Tutukov}(1984)}]{1984ApJS...54..335I}
{Iben}, Jr. I., {Tutukov} A.~V., 1984, \apjs, 54, 335

\bibitem[{{Johansson} {et~al}\mbox{.}(2013){Johansson}, {Thomas}, {Pforr},
  {Maraston}, {Nichol}, {Smith}, {Lampeitl}, {Beifiori}, {Gupta}, \&
  {Schneider}}]{2013MNRAS.435.1680J}
{Johansson} J. {et~al.}, 2013, \mnras, 435, 1680

\bibitem[{{Kelly}(2007)}]{2007ApJ...665.1489K}
{Kelly} B.~C., 2007, \apj, 665, 1489

\bibitem[{{Kelly} {et~al}\mbox{.}(2010){Kelly}, {Hicken}, {Burke}, {Mandel}, \&
  {Kirshner}}]{2010ApJ...715..743K}
{Kelly} P.~L., {Hicken} M., {Burke} D.~L., {Mandel} K.~S., {Kirshner} R.~P.,
  2010, \apj, 715, 743

\bibitem[{{Kennicutt}(1998)}]{1998ARA&A..36..189K}
{Kennicutt} J. R.~C., 1998, \araa, 36, 189

\bibitem[{{Kessler} {et~al}\mbox{.}(2009){Kessler}, {Becker}, {Cinabro},
  {Vanderplas}, {Frieman}, {Marriner}, {Davis}, {Dilday}, {Holtzman}, {Jha},
  {Lampeitl}, {Sako}, {Smith}, {Zheng}, {Nichol}, {Bassett}, {Bender}, {Depoy},
  {Doi}, {Elson}, {Filippenko}, {Foley}, {Garnavich}, {Hopp}, {Ihara},
  {Ketzeback}, {Kollatschny}, {Konishi}, {Marshall}, {McMillan}, {Miknaitis},
  {Morokuma}, {M{\"o}rtsell}, {Pan}, {Prieto}, {Richmond}, {Riess}, {Romani},
  {Schneider}, {Sollerman}, {Takanashi}, {Tokita}, {van der Heyden}, {Wheeler},
  {Yasuda}, \& {York}}]{2009ApJS..185...32K}
{Kessler} R. {et~al.}, 2009, \apjs, 185, 32

\bibitem[{{Kewley} {et~al}\mbox{.}(2001){Kewley}, {Dopita}, {Sutherland},
  {Heisler}, \& {Trevena}}]{2001ApJ...556..121K}
{Kewley} L.~J., {Dopita} M.~A., {Sutherland} R.~S., {Heisler} C.~A., {Trevena}
  J., 2001, \apj, 556, 121

\bibitem[{{Kewley} \& {Ellison}(2008)}]{2008ApJ...681.1183K}
{Kewley} L.~J., {Ellison} S.~L., 2008, \apj, 681, 1183

\bibitem[{{Lamareille} {et~al}\mbox{.}(2006){Lamareille}, {Contini}, {Le
  Borgne}, {Brinchmann}, {Charlot}, \& {Richard}}]{2006A&A...448..893L}
{Lamareille} F., {Contini} T., {Le Borgne} J.-F., {Brinchmann} J., {Charlot}
  S., {Richard} J., 2006, \aap, 448, 893

\bibitem[{{Lampeitl} {et~al}\mbox{.}(2010){Lampeitl}, {Smith}, {Nichol},
  {Bassett}, {Cinabro}, {Dilday}, {Foley}, {Frieman}, {Garnavich}, {Goobar},
  {Im}, {Jha}, {Marriner}, {Miquel}, {Nordin}, {{\"O}stman}, {Riess}, {Sako},
  {Schneider}, {Sollerman}, \& {Stritzinger}}]{2010ApJ...722..566L}
{Lampeitl} H. {et~al.}, 2010, \apj, 722, 566

\bibitem[{{Lantz} {et~al}\mbox{.}(2004){Lantz}, {Aldering}, {Antilogus},
  {Bonnaud}, {Capoani}, {Castera}, {Copin}, {Dubet}, {Gangler}, {Henault},
  {Lemonnier}, {Pain}, {Pecontal}, {Pecontal}, \&
  {Smadja}}]{2004SPIE.5249..146L}
{Lantz} B. {et~al.}, 2004, in Society of Photo-Optical Instrumentation
  Engineers (SPIE) Conference Series, Vol. 5249, Optical Design and
  Engineering, {Mazuray} L., {Rogers} P.~J., {Wartmann} R., eds., pp. 146--155

\bibitem[{{Law} {et~al}\mbox{.}(2009){Law}, {Kulkarni}, {Dekany}, {Ofek},
  {Quimby}, {Nugent}, {Surace}, {Grillmair}, {Bloom}, {Kasliwal}, {Bildsten},
  {Brown}, {Cenko}, {Ciardi}, {Croner}, {Djorgovski}, {van Eyken},
  {Filippenko}, {Fox}, {Gal-Yam}, {Hale}, {Hamam}, {Helou}, {Henning},
  {Howell}, {Jacobsen}, {Laher}, {Mattingly}, {McKenna}, {Pickles},
  {Poznanski}, {Rahmer}, {Rau}, {Rosing}, {Shara}, {Smith}, {Starr},
  {Sullivan}, {Velur}, {Walters}, \& {Zolkower}}]{2009PASP..121.1395L}
{Law} N.~M. {et~al.}, 2009, \pasp, 121, 1395

\bibitem[{{Le Borgne} \& {Rocca-Volmerange}(2002)}]{2002A&A...386..446L}
{Le Borgne} D., {Rocca-Volmerange} B., 2002, \aap, 386, 446

\bibitem[{{Lentz} {et~al}\mbox{.}(2000){Lentz}, {Baron}, {Branch},
  {Hauschildt}, \& {Nugent}}]{2000ApJ...530..966L}
{Lentz} E.~J., {Baron} E., {Branch} D., {Hauschildt} P.~H., {Nugent} P.~E.,
  2000, \apj, 530, 966

\bibitem[{{Maguire} {et~al}\mbox{.}(2012){Maguire}, {Sullivan}, {Ellis},
  {Nugent}, {Howell}, {Gal-Yam}, {Cooke}, {Mazzali}, {Pan}, {Dilday}, {Thomas},
  {Arcavi}, {Ben-Ami}, {Bersier}, {Bianco}, {Fulton}, {Hook}, {Horesh},
  {Hsiao}, {James}, {Podsiadlowski}, {Walker}, {Yaron}, {Kasliwal}, {Laher},
  {Law}, {Ofek}, {Poznanski}, \& {Surace}}]{2012MNRAS.426.2359M}
{Maguire} K. {et~al.}, 2012, \mnras, 426, 2359

\bibitem[{{Maguire} {et~al}\mbox{.}(2014){Maguire}, {Sullivan}, {Pan},
  {Gal-Yam}, {Hook}, {Howell}, {Nugent}, {Mazzali}, {Chotard}, {Clubb},
  {Filippenko}, {Kasliwal}, {Kandrashoff}, {Poznanski}, {Saunders},
  {Silverman}, {Walker}, \& {Xu}}]{2014MNRAS.444.3258M}
{Maguire} K. {et~al.}, 2014, \mnras, 444, 3258

\bibitem[{{Maguire} {et~al}\mbox{.}(2013){Maguire}, {Sullivan}, {Patat},
  {Gal-Yam}, {Hook}, {Dhawan}, {Howell}, {Mazzali}, {Nugent}, {Pan},
  {Podsiadlowski}, {Simon}, {Sternberg}, {Valenti}, {Baltay}, {Bersier},
  {Blagorodnova}, {Chen}, {Ellman}, {Feindt}, {F{\"o}rster}, {Fraser},
  {Gonz{\'a}lez-Gait{\'a}n}, {Graham}, {Guti{\'e}rrez}, {Hachinger},
  {Hadjiyska}, {Inserra}, {Knapic}, {Laher}, {Leloudas}, {Margheim},
  {McKinnon}, {Molinaro}, {Morrell}, {Ofek}, {Rabinowitz}, {Rest}, {Sand},
  {Smareglia}, {Smartt}, {Taddia}, {Walker}, {Walton}, \&
  {Young}}]{2013MNRAS.436..222M}
{Maguire} K. {et~al.}, 2013, \mnras, 436, 222

\bibitem[{{Mannucci} {et~al}\mbox{.}(2005){Mannucci}, {Della Valle}, {Panagia},
  {Cappellaro}, {Cresci}, {Maiolino}, {Petrosian}, \&
  {Turatto}}]{2005A&A...433..807M}
{Mannucci} F., {Della Valle} M., {Panagia} N., {Cappellaro} E., {Cresci} G.,
  {Maiolino} R., {Petrosian} A., {Turatto} M., 2005, \aap, 433, 807

\bibitem[{{Maoz} {et~al}\mbox{.}(2013){Maoz}, {Mannucci}, \&
  {Nelemans}}]{2013arXiv1312.0628M}
{Maoz} D., {Mannucci} F., {Nelemans} G., 2013, ArXiv e-prints

\bibitem[{{Marion} {et~al}\mbox{.}(2013){Marion}, {Vinko}, {Wheeler}, {Foley},
  {Hsiao}, {Brown}, {Challis}, {Filippenko}, {Garnavich}, {Kirshner},
  {Landsman}, {Parrent}, {Pritchard}, {Roming}, {Silverman}, \&
  {Wang}}]{2013ApJ...777...40M}
{Marion} G.~H. {et~al.}, 2013, \apj, 777, 40

\bibitem[{{Markwardt}(2009)}]{2009ASPC..411..251M}
{Markwardt} C.~B., 2009, in Astronomical Society of the Pacific Conference
  Series, Vol. 411, Astronomical Data Analysis Software and Systems XVIII,
  {Bohlender} D.~A., {Durand} D., {Dowler} P., eds., p. 251

\bibitem[{{Mazzali} {et~al}\mbox{.}(2005{\natexlab{a}}){Mazzali}, {Benetti},
  {Altavilla}, {Blanc}, {Cappellaro}, {Elias-Rosa}, {Garavini}, {Goobar},
  {Harutyunyan}, {Kotak}, {Leibundgut}, {Lundqvist}, {Mattila}, {Mendez},
  {Nobili}, {Pain}, {Pastorello}, {Patat}, {Pignata}, {Podsiadlowski},
  {Ruiz-Lapuente}, {Salvo}, {Schmidt}, {Sollerman}, {Stanishev}, {Stehle},
  {Tout}, {Turatto}, \& {Hillebrandt}}]{2005ApJ...623L..37M}
{Mazzali} P.~A. {et~al.}, 2005{\natexlab{a}}, \apjl, 623, L37

\bibitem[{{Mazzali} {et~al}\mbox{.}(2005{\natexlab{b}}){Mazzali}, {Benetti},
  {Stehle}, {Branch}, {Deng}, {Maeda}, {Nomoto}, \&
  {Hamuy}}]{2005MNRAS.357..200M}
{Mazzali} P.~A., {Benetti} S., {Stehle} M., {Branch} D., {Deng} J., {Maeda} K.,
  {Nomoto} K., {Hamuy} M., 2005{\natexlab{b}}, \mnras, 357, 200

\bibitem[{{Mazzali} {et~al}\mbox{.}(2007){Mazzali}, {R{\"o}pke}, {Benetti}, \&
  {Hillebrandt}}]{2007Sci...315..825M}
{Mazzali} P.~A., {R{\"o}pke} F.~K., {Benetti} S., {Hillebrandt} W., 2007,
  Science, 315, 825

\bibitem[{{Miller} \& {Stone}(1993)}]{Kast_spectrograph}
{Miller} J.~S., {Stone} R. P.~S., 1993, Lick Obs. Tech. Rep., No. 66

\bibitem[{{Nugent} {et~al}\mbox{.}(1995){Nugent}, {Phillips}, {Baron},
  {Branch}, \& {Hauschildt}}]{1995ApJ...455L.147N}
{Nugent} P., {Phillips} M., {Baron} E., {Branch} D., {Hauschildt} P., 1995,
  \apjl, 455, L147

\bibitem[{{Nugent} {et~al}\mbox{.}(2011){Nugent}, {Sullivan}, {Cenko},
  {Thomas}, {Kasen}, {Howell}, {Bersier}, {Bloom}, {Kulkarni}, {Kandrashoff},
  {Filippenko}, {Silverman}, {Marcy}, {Howard}, {Isaacson}, {Maguire},
  {Suzuki}, {Tarlton}, {Pan}, {Bildsten}, {Fulton}, {Parrent}, {Sand},
  {Podsiadlowski}, {Bianco}, {Dilday}, {Graham}, {Lyman}, {James}, {Kasliwal},
  {Law}, {Quimby}, {Hook}, {Walker}, {Mazzali}, {Pian}, {Ofek}, {Gal-Yam}, \&
  {Poznanski}}]{2011Natur.480..344N}
{Nugent} P.~E. {et~al.}, 2011, \nat, 480, 344

\bibitem[{{Oke} {et~al}\mbox{.}(1995){Oke}, {Cohen}, {Carr}, {Cromer},
  {Dingizian}, {Harris}, {Labrecque}, {Lucinio}, {Schaal}, {Epps}, \&
  {Miller}}]{1995PASP..107..375O}
{Oke} J.~B. {et~al.}, 1995, \pasp, 107, 375

\bibitem[{{Oke} \& {Gunn}(1982)}]{1982PASP...94..586O}
{Oke} J.~B., {Gunn} J.~E., 1982, \pasp, 94, 586

\bibitem[{{Pan} {et~al}\mbox{.}(2014){Pan}, {Sullivan}, {Maguire}, {Hook},
  {Nugent}, {Howell}, {Arcavi}, {Botyanszki}, {Cenko}, {DeRose}, {Fakhouri},
  {Gal-Yam}, {Hsiao}, {Kulkarni}, {Laher}, {Lidman}, {Nordin}, {Walker}, \&
  {Xu}}]{2014MNRAS.438.1391P}
{Pan} Y.-C. {et~al.}, 2014, \mnras, 438, 1391

\bibitem[{{Patat} {et~al}\mbox{.}(2009){Patat}, {Baade}, {H{\"o}flich},
  {Maund}, {Wang}, \& {Wheeler}}]{2009A&A...508..229P}
{Patat} F., {Baade} D., {H{\"o}flich} P., {Maund} J.~R., {Wang} L., {Wheeler}
  J.~C., 2009, \aap, 508, 229

\bibitem[{{Perlmutter} {et~al}\mbox{.}(1999){Perlmutter}, {Aldering},
  {Goldhaber}, {Knop}, {Nugent}, {Castro}, {Deustua}, {Fabbro}, {Goobar},
  {Groom}, {Hook}, {Kim}, {Kim}, {Lee}, {Nunes}, {Pain}, {Pennypacker},
  {Quimby}, {Lidman}, {Ellis}, {Irwin}, {McMahon}, {Ruiz-Lapuente}, {Walton},
  {Schaefer}, {Boyle}, {Filippenko}, {Matheson}, {Fruchter}, {Panagia},
  {Newberg}, {Couch}, \& {The Supernova Cosmology
  Project}}]{1999ApJ...517..565P}
{Perlmutter} S. {et~al.}, 1999, \apj, 517, 565

\bibitem[{{Pettini} \& {Pagel}(2004)}]{2004MNRAS.348L..59P}
{Pettini} M., {Pagel} B.~E.~J., 2004, \mnras, 348, L59

\bibitem[{{Quimby} {et~al}\mbox{.}(2006){Quimby}, {H{\"o}flich}, {Kannappan},
  {Rykoff}, {Rujopakarn}, {Akerlof}, {Gerardy}, \&
  {Wheeler}}]{2006ApJ...636..400Q}
{Quimby} R., {H{\"o}flich} P., {Kannappan} S.~J., {Rykoff} E., {Rujopakarn} W.,
  {Akerlof} C.~W., {Gerardy} C.~L., {Wheeler} J.~C., 2006, \apj, 636, 400

\bibitem[{{Rahmer} {et~al}\mbox{.}(2008){Rahmer}, {Smith}, {Velur}, {Hale},
  {Law}, {Bui}, {Petrie}, \& {Dekany}}]{2008SPIE.7014E.163R}
{Rahmer} G., {Smith} R., {Velur} V., {Hale} D., {Law} N., {Bui} K., {Petrie}
  H., {Dekany} R., 2008, in Society of Photo-Optical Instrumentation Engineers
  (SPIE) Conference Series, Vol. 7014, Society of Photo-Optical Instrumentation
  Engineers (SPIE) Conference Series

\bibitem[{{Rau} {et~al}\mbox{.}(2009){Rau}, {Kulkarni}, {Law}, {Bloom},
  {Ciardi}, {Djorgovski}, {Fox}, {Gal-Yam}, {Grillmair}, {Kasliwal}, {Nugent},
  {Ofek}, {Quimby}, {Reach}, {Shara}, {Bildsten}, {Cenko}, {Drake},
  {Filippenko}, {Helfand}, {Helou}, {Howell}, {Poznanski}, \&
  {Sullivan}}]{2009PASP..121.1334R}
{Rau} A. {et~al.}, 2009, \pasp, 121, 1334

\bibitem[{{Rest} {et~al}\mbox{.}(2013){Rest}, {Scolnic}, {Foley}, {Huber},
  {Chornock}, {Narayan}, {Tonry}, {Berger}, {Soderberg}, {Stubbs}, {Riess},
  {Kirshner}, {Smartt}, {Schlafly}, {Rodney}, {Botticella}, {Brout}, {Challis},
  {Czekala}, {Drout}, {Hudson}, {Kotak}, {Leibler}, {Lunnan}, {Marion},
  {McCrum}, {Milisavljevic}, {Pastorello}, {Sanders}, {Smith}, {Stafford},
  {Thilker}, {Valenti}, {Wood-Vasey}, {Zheng}, {Burgett}, {Chambers},
  {Denneau}, {Draper}, {Flewelling}, {Hodapp}, {Kaiser}, {Kudritzki},
  {Magnier}, {Metcalfe}, {Price}, {Sweeney}, {Wainscoat}, \&
  {Waters}}]{2013arXiv1310.3828R}
{Rest} A. {et~al.}, 2013, ArXiv e-prints

\bibitem[{{Riess} {et~al}\mbox{.}(1998){Riess}, {Filippenko}, {Challis},
  {Clocchiatti}, {Diercks}, {Garnavich}, {Gilliland}, {Hogan}, {Jha},
  {Kirshner}, {Leibundgut}, {Phillips}, {Reiss}, {Schmidt}, {Schommer},
  {Smith}, {Spyromilio}, {Stubbs}, {Suntzeff}, \&
  {Tonry}}]{1998AJ....116.1009R}
{Riess} A.~G. {et~al.}, 1998, \aj, 116, 1009

\bibitem[{{Riess} {et~al}\mbox{.}(2007){Riess}, {Strolger}, {Casertano},
  {Ferguson}, {Mobasher}, {Gold}, {Challis}, {Filippenko}, {Jha}, {Li},
  {Tonry}, {Foley}, {Kirshner}, {Dickinson}, {MacDonald}, {Eisenstein},
  {Livio}, {Younger}, {Xu}, {Dahl{\'e}n}, \& {Stern}}]{2007ApJ...659...98R}
{Riess} A.~G. {et~al.}, 2007, \apj, 659, 98

\bibitem[{{Rigault} {et~al}\mbox{.}(2013){Rigault}, {Copin}, {Aldering},
  {Antilogus}, {Aragon}, {Bailey}, {Baltay}, {Bongard}, {Buton}, {Canto},
  {Cellier-Holzem}, {Childress}, {Chotard}, {Fakhouri}, {Feindt}, {Fleury},
  {Gangler}, {Greskovic}, {Guy}, {Kim}, {Kowalski}, {Lombardo}, {Nordin},
  {Nugent}, {Pain}, {P{\'e}contal}, {Pereira}, {Perlmutter}, {Rabinowitz},
  {Runge}, {Saunders}, {Scalzo}, {Smadja}, {Tao}, {Thomas}, \&
  {Weaver}}]{2013A&A...560A..66R}
{Rigault} M. {et~al.}, 2013, \aap, 560, A66

\bibitem[{{Salpeter}(1955)}]{1955ApJ...121..161S}
{Salpeter} E.~E., 1955, \apj, 121, 161

\bibitem[{{S{\'a}nchez-Bl{\'a}zquez}
  {et~al}\mbox{.}(2006){S{\'a}nchez-Bl{\'a}zquez}, {Peletier},
  {Jim{\'e}nez-Vicente}, {Cardiel}, {Cenarro}, {Falc{\'o}n-Barroso}, {Gorgas},
  {Selam}, \& {Vazdekis}}]{2006MNRAS.371..703S}
{S{\'a}nchez-Bl{\'a}zquez} P. {et~al.}, 2006, \mnras, 371, 703

\bibitem[{{Sarzi} {et~al}\mbox{.}(2006){Sarzi}, {Falc{\'o}n-Barroso}, {Davies},
  {Bacon}, {Bureau}, {Cappellari}, {de Zeeuw}, {Emsellem}, {Fathi},
  {Krajnovi{\'c}}, {Kuntschner}, {McDermid}, \&
  {Peletier}}]{2006MNRAS.366.1151S}
{Sarzi} M. {et~al.}, 2006, \mnras, 366, 1151

\bibitem[{{Schaefer} \& {Pagnotta}(2012)}]{2012Natur.481..164S}
{Schaefer} B.~E., {Pagnotta} A., 2012, \nat, 481, 164

\bibitem[{{Silverman} {et~al}\mbox{.}(2012{\natexlab{a}}){Silverman},
  {Ganeshalingam}, {Li}, \& {Filippenko}}]{2012MNRAS.425.1889S}
{Silverman} J.~M., {Ganeshalingam} M., {Li} W., {Filippenko} A.~V.,
  2012{\natexlab{a}}, \mnras, 425, 1889

\bibitem[{{Silverman} {et~al}\mbox{.}(2012{\natexlab{b}}){Silverman}, {Kong},
  \& {Filippenko}}]{2012MNRAS.425.1819S}
{Silverman} J.~M., {Kong} J.~J., {Filippenko} A.~V., 2012{\natexlab{b}},
  \mnras, 425, 1819

\bibitem[{{Smith} {et~al}\mbox{.}(2011){Smith}, {Lynn}, {Sullivan}, {Lintott},
  {Nugent}, {Botyanszki}, {Kasliwal}, {Quimby}, {Bamford}, {Fortson},
  {Schawinski}, {Hook}, {Blake}, {Podsiadlowski}, {J{\"o}nsson}, {Gal-Yam},
  {Arcavi}, {Howell}, {Bloom}, {Jacobsen}, {Kulkarni}, {Law}, {Ofek}, \&
  {Walters}}]{2011MNRAS.412.1309S}
{Smith} A.~M. {et~al.}, 2011, \mnras, 412, 1309

\bibitem[{{Steele} {et~al}\mbox{.}(2004){Steele}, {Smith}, {Rees}, {Baker},
  {Bates}, {Bode}, {Bowman}, {Carter}, {Etherton}, {Ford}, {Fraser}, {Gomboc},
  {Lett}, {Mansfield}, {Marchant}, {Medrano-Cerda}, {Mottram}, {Raback},
  {Scott}, {Tomlinson}, \& {Zamanov}}]{2004SPIE.5489..679S}
{Steele} I.~A. {et~al.}, 2004, in Society of Photo-Optical Instrumentation
  Engineers (SPIE) Conference Series, Vol. 5489, Society of Photo-Optical
  Instrumentation Engineers (SPIE) Conference Series, {Oschmann} Jr. J.~M.,
  ed., pp. 679--692

\bibitem[{{Sternberg} {et~al}\mbox{.}(2011){Sternberg}, {Gal-Yam}, {Simon},
  {Leonard}, {Quimby}, {Phillips}, {Morrell}, {Thompson}, {Ivans}, {Marshall},
  {Filippenko}, {Marcy}, {Bloom}, {Patat}, {Foley}, {Yong}, {Penprase},
  {Beeler}, {Allende Prieto}, \& {Stringfellow}}]{2011Sci...333..856S}
{Sternberg} A. {et~al.}, 2011, Science, 333, 856

\bibitem[{{Strateva} {et~al}\mbox{.}(2001){Strateva}, {Ivezi{\'c}}, {Knapp},
  {Narayanan}, {Strauss}, {Gunn}, {Lupton}, {Schlegel}, {Bahcall}, {Brinkmann},
  {Brunner}, {Budav{\'a}ri}, {Csabai}, {Castander}, {Doi}, {Fukugita}, {Gy{\H
  o}ry}, {Hamabe}, {Hennessy}, {Ichikawa}, {Kunszt}, {Lamb}, {McKay},
  {Okamura}, {Racusin}, {Sekiguchi}, {Schneider}, {Shimasaku}, \&
  {York}}]{2001AJ....122.1861S}
{Strateva} I. {et~al.}, 2001, \aj, 122, 1861

\bibitem[{{Sullivan} {et~al}\mbox{.}(2010){Sullivan}, {Conley}, {Howell},
  {Neill}, {Astier}, {Balland}, {Basa}, {Carlberg}, {Fouchez}, {Guy}, {Hardin},
  {Hook}, {Pain}, {Palanque-Delabrouille}, {Perrett}, {Pritchet}, {Regnault},
  {Rich}, {Ruhlmann-Kleider}, {Baumont}, {Hsiao}, {Kronborg}, {Lidman},
  {Perlmutter}, \& {Walker}}]{2010MNRAS.406..782S}
{Sullivan} M. {et~al.}, 2010, \mnras, 406, 782

\bibitem[{{Sullivan} {et~al}\mbox{.}(2006){Sullivan}, {Le Borgne}, {Pritchet},
  {Hodsman}, {Neill}, {Howell}, {Carlberg}, {Astier}, {Aubourg}, {Balam},
  {Basa}, {Conley}, {Fabbro}, {Fouchez}, {Guy}, {Hook}, {Pain},
  {Palanque-Delabrouille}, {Perrett}, {Regnault}, {Rich}, {Taillet}, {Baumont},
  {Bronder}, {Ellis}, {Filiol}, {Lusset}, {Perlmutter}, {Ripoche}, \&
  {Tao}}]{2006ApJ...648..868S}
{Sullivan} M. {et~al.}, 2006, \apj, 648, 868

\bibitem[{{Sullivan} {et~al}\mbox{.}(2011){Sullivan}
  {et~al.}}]{2011ApJ...737..102S}
{Sullivan} M., {et~al.}, 2011, \apj, 737, 102

\bibitem[{{Tanaka} {et~al}\mbox{.}(2008){Tanaka}, {Mazzali}, {Benetti},
  {Nomoto}, {Elias-Rosa}, {Kotak}, {Pignata}, {Stanishev}, \&
  {Hachinger}}]{2008ApJ...677..448T}
{Tanaka} M. {et~al.}, 2008, \apj, 677, 448

\bibitem[{{Tanaka} {et~al}\mbox{.}(2006){Tanaka}, {Mazzali}, {Maeda}, \&
  {Nomoto}}]{2006ApJ...645..470T}
{Tanaka} M., {Mazzali} P.~A., {Maeda} K., {Nomoto} K., 2006, \apj, 645, 470

\bibitem[{{Tremonti} {et~al}\mbox{.}(2004){Tremonti}, {Heckman}, {Kauffmann},
  {Brinchmann}, {Charlot}, {White}, {Seibert}, {Peng}, {Schlegel}, {Uomoto},
  {Fukugita}, \& {Brinkmann}}]{2004ApJ...613..898T}
{Tremonti} C.~A. {et~al.}, 2004, \apj, 613, 898

\bibitem[{{Vazdekis} {et~al}\mbox{.}(2010){Vazdekis},
  {S{\'a}nchez-Bl{\'a}zquez}, {Falc{\'o}n-Barroso}, {Cenarro}, {Beasley},
  {Cardiel}, {Gorgas}, \& {Peletier}}]{2010MNRAS.404.1639V}
{Vazdekis} A., {S{\'a}nchez-Bl{\'a}zquez} P., {Falc{\'o}n-Barroso} J.,
  {Cenarro} A.~J., {Beasley} M.~A., {Cardiel} N., {Gorgas} J., {Peletier}
  R.~F., 2010, \mnras, 404, 1639

\bibitem[{{Vernet} {et~al}\mbox{.}(2011){Vernet}, {Dekker}, {D'Odorico},
  {Kaper}, {Kjaergaard}, {Hammer}, {Randich}, {Zerbi}, {Groot}, {Hjorth},
  {Guinouard}, {Navarro}, {Adolfse}, {Albers}, {Amans}, {Andersen}, {Andersen},
  {Binetruy}, {Bristow}, {Castillo}, {Chemla}, {Christensen}, {Conconi},
  {Conzelmann}, {Dam}, {de Caprio}, {de Ugarte Postigo}, {Delabre}, {di
  Marcantonio}, {Downing}, {Elswijk}, {Finger}, {Fischer}, {Flores}, {Fran{\c
  c}ois}, {Goldoni}, {Guglielmi}, {Haigron}, {Hanenburg}, {Hendriks},
  {Horrobin}, {Horville}, {Jessen}, {Kerber}, {Kern}, {Kiekebusch}, {Kleszcz},
  {Klougart}, {Kragt}, {Larsen}, {Lizon}, {Lucuix}, {Mainieri}, {Manuputy},
  {Martayan}, {Mason}, {Mazzoleni}, {Michaelsen}, {Modigliani}, {Moehler},
  {M{\o}ller}, {Norup S{\o}rensen}, {N{\o}rregaard}, {P{\'e}roux}, {Patat},
  {Pena}, {Pragt}, {Reinero}, {Rigal}, {Riva}, {Roelfsema}, {Royer}, {Sacco},
  {Santin}, {Schoenmaker}, {Spano}, {Sweers}, {Ter Horst}, {Tintori}, {Tromp},
  {van Dael}, {van der Vliet}, {Venema}, {Vidali}, {Vinther}, {Vola},
  {Winters}, {Wistisen}, {Wulterkens}, \& {Zacchei}}]{2011A&A...536A.105V}
{Vernet} J. {et~al.}, 2011, \aap, 536, A105

\bibitem[{{Wang} {et~al}\mbox{.}(2003){Wang}, {Baade}, {H{\"o}flich},
  {Khokhlov}, {Wheeler}, {Kasen}, {Nugent}, {Perlmutter}, {Fransson}, \&
  {Lundqvist}}]{2003ApJ...591.1110W}
{Wang} L. {et~al.}, 2003, \apj, 591, 1110

\bibitem[{{Wang} {et~al}\mbox{.}(2009){Wang}, {Filippenko}, {Ganeshalingam},
  {Li}, {Silverman}, {Wang}, {Chornock}, {Foley}, {Gates}, {Macomber},
  {Serduke}, {Steele}, \& {Wong}}]{2009ApJ...699L.139W}
{Wang} X. {et~al.}, 2009, \apjl, 699, L139

\bibitem[{{Wang} {et~al}\mbox{.}(2013){Wang}, {Wang}, {Filippenko}, {Zhang}, \&
  {Zhao}}]{2013Sci...340..170W}
{Wang} X., {Wang} L., {Filippenko} A.~V., {Zhang} T., {Zhao} X., 2013, Science,
  340, 170

\bibitem[{{Webbink}(1984)}]{1984ApJ...277..355W}
{Webbink} R.~F., 1984, \apj, 277, 355

\bibitem[{{Whelan} \& {Iben}(1973)}]{1973ApJ...186.1007W}
{Whelan} J., {Iben}, Jr. I., 1973, \apj, 186, 1007

\bibitem[{{Yaron} \& {Gal-Yam}(2012)}]{2012PASP..124..668Y}
{Yaron} O., {Gal-Yam} A., 2012, \pasp, 124, 668

\end{thebibliography}

\label{lastpage}

\end{document}